\documentclass{aa}  
\usepackage{dcolumn}
\usepackage{graphicx}
\usepackage{txfonts}
\newcommand{\Nl}[3]{#1\,{\sc #2}\,$\lambda{#3}$}
\makeatletter
 \def\hlinewd#1{%
   \noalign{\ifnum0=`}\fi\hrule \@height #1 \futurelet
    \reserved@a\@xhline}
\makeatother
\newcommand{\htopline}{\hlinewd{.8pt}}
\newcommand{\hmidline}{\hlinewd{.2pt}}
\newcommand{\hbotline}{\htopline}
\newcommand{\mcc}[1]{\multicolumn{1}{c}{#1}}

\newcommand{\kms}{km\,s$^{-1}$}
\newcolumntype{d}{D{.}{.}{-1}}

\begin{document}

    \title{NGC\,6240: A triple nucleus
    system in the advanced or final state of merging
}

   \author{W. Kollatschny \inst{1}, 
           P. M. Weilbacher \inst{2},
           M. W. Ochmann \inst{1},
           D. Chelouche \inst{3}, 
           A. Monreal-Ibero \inst{4,5}, 
           R. Bacon \inst{6},
           T. Contini \inst{7}
          }

   \institute{Institut f\"ur Astrophysik, Universit\"at G\"ottingen,
              Friedrich-Hund Platz 1, D-37077 G\"ottingen, Germany\\
              \email{wkollat@astro.physik.uni-goettingen.de}
         \and
          Leibniz-Institut f\"ur Astrophysik Potsdam (AIP),
          An der Sternwarte 16, D-14482 Potsdam, Germany
         \and
   Physics Department and the Haifa Research Center for Theoretical Physics and
   Astrophysics, University of Haifa, Haifa 3498838, Israel
         \and         
          Instituto de Astrof\'{\i}sica de Canarias (IAC), E-38205 La Laguna, Tenerife, Spain
         \and
         Universidad de La Laguna, Dpto.\ Astrof\'{\i}sica, E-38206 La Laguna, Tenerife, Spain
         \and           
    Univ Lyon, Univ Lyon1, Ens de Lyon, CNRS, Centre de Recherche Astrophysique
    de Lyon UMR5574, F-69230, SAint-Genis-Laval, France     
         \and  
         Institut de Recherche en Astrophysique et Plan\'etologie (IRAP),
        Universit\'e de Toulouse, CNRS, UPS, F-31400 Toulouse, France
         }

   \date{Received 21 August 2019; Accepted 27 October 2019}
   \authorrunning{Kollatschny et al.}
   \titlerunning{NGC\,6240: a triple nucleus system}
\abstract{}{}{}{}{} 
% 5 {} token are mandatory
 
\abstract
% context heading (optional)
 {}
 { NGC\,6240 is a well-studied nearby galaxy system in the process of merging. Based on optical, X-ray, and radio observations, it is thought 
   to harbor two active nuclei.
 We carried out a detailed optical 3D spectroscopic study to investigate the inner region of 
 this system in connection with existing MERLIN and VLBA data.} 
 % methods heading (mandatory) 
 {We observed NGC\,6240 with very high spatial resolution using the MUSE instrument in the Narrow-Field Mode
 with the four-laser GALACSI adaptive optics system on the ESO VLT under seeing conditions of $0\farcs49$. Our 3D spectra cover 
 the wavelength range from 4725 to 9350\,\AA{} at a spatial resolution of $\sim75$ mas.}
  % conclusions heading (optional), leave it empty if necessary 
  {We report the discovery of three nuclei in the final state of merging within a region of only 1 kpc in the NGC\,6240 system.
  Thanks to MUSE we are able to show that the formerly unresolved southern component actually consists of two 
  distinct nuclei separated by only 198 pc. In combination with Gaia data we reach an 
  absolute positional accuracy of only 30\,mas that is essential to compare optical spectra 
  with MERLIN and VLBA radio positions.}
 { The verification and detailed study of a system  with three nuclei, two of which are active and each 
 with a mass in excess of $9\times10^{7} M_{\odot}$,  is of great importance
  for the understanding of hierarchical galaxy formation via merging processes since multiple mergers lead to a 
  faster evolution of massive galaxies in comparison to binary mergers. So far it has been suggested that the formation of 
  galactic nuclei with multiple supermassive black holes (SMBHs) is expected to be rare in the local universe.
  Triple massive black hole systems might be of fundamental importance for the coalescence of massive black hole binaries 
  in less than a Hubble time leading to the loudest sources of gravitational waves in the millihertz regime.}


 \maketitle
%
%________________________________________________________________

\section{Introduction}

It is generally accepted that all massive galaxies host supermassive
black holes (SMBHs) in their centers (Kormendy \& Ho\citealt{kormendy13}) 
and that mergers of two galaxies lead to the formation of black hole binaries.
These binaries might evolve into
single or double active galactic nuclei (AGN)
if their nuclei are accreting gas (Begelman et al.\citealt{begelman80}).
If the binary lifetime exceeds the typical time between mergers, triple black hole systems
may form (Hoffman \& Loeb \citealt{hoffman07}).
 Of these systems, the most interesting ones  are those in an advanced state of merging (i.e., those with the smallest distances between their SMBHs).  
The closest known optical or infrared distances of two nuclei correspond to
a projected separation of $\sim$\,one kiloparsec, except for only one example, MCG+02-21-013, with a projected separation of 300 pc 
(see Table 1 in Koss et al.\citealt{koss18}). 
Arp\,220, which is the nearest ultraluminous infrared galaxy (ULIRG), 
has two near-IR nuclei at a separation of 330 pc (Genzel et al.\citealt{genzel01},
and references therein). The detection of another close-separation binary quasar 
has been reported recently (Goulding et al.\citealt{goulding19}). The nuclei in 
SDSS J1010+1413 are separated by 430 pc only.

Here we present  high spatial resolution observations of
the nearby merging galaxy system NGC\,6240 
obtained with the MUSE instrument at the ESO VLT.
NGC\,6240 is one of the nearest ULIRGs
(Genzel \& Cesarsky\citealt{genzel00}).
It is at the faint limit of the ULIRG class with respect to its far-IR luminosity 
(Wright et al.\citealt{wright84}).
NGC\,6240  belongs to the class of merging systems where the galaxies are
separated by less than 10 kpc based on optical and IR images
(Koss et al.\citealt{koss18}).
It is the merging system with the second smallest separation
between the two components in the list of Koss et al.\cite{koss18},
corresponding to a projected separation of 900 pc between its
northern and southern components. 
It has been proposed that NGC\,6240 is a merger of two massive
disk galaxies (e.g., Fosbury \& Wall \citealt{fosbury79}, Engel et al.
\citealt{engel10} and references therein).\\
NGC\,6240 has been the subject of numerous studies.
Based on radio observations with MERLIN and VLBA  
(Gallimore et al.~\citealt{gallimore04}) and 
X-ray observations with Chandra (Komossa et al.\citealt{komossa03}),
it has been claimed that NGC\,6240 hosts a pair of AGN.
However, to date the exact location of the two active nuclei is not  accurately known 
as the two radio positions
are separated by 1.51 arcsec (e.g., Max et al.\citealt{max07}),
while the two brightest near-IR or optical spots are separated by 1.8 arcsec.\\
There are very few other cases of merging systems where the two
nuclei are separated by less than 10 kpc and at the same time show Seyfert characteristics
(e.g., Satyapal et al.\citealt{satyapal17}, their Table 8). Among those double
nucleus Seyfert galaxies are objects 
like  Mrk\,266 (Kollatschny et al.\citealt{kollatschny84,kollatschny98};
Mazzarella et al.\citealt{mazzarella12}) and Mrk\,739 (Netzer et al.\citealt{netzer87})
that have been studied in detail for many years.

NGC\,6240 has a mean redshift of z = 0.02448 (Downes et al.\citealt{downes93})
corresponding to 7339~\kms.
Throughout this paper we assume $\Lambda$CDM cosmology with a Hubble constant
of H$_0$~=~73~\kms Mpc$^{-1}$, $\Omega_{\rm M}$=0.27, and $\Omega_{\Lambda}$=0.73.
Following the cosmological calculator by Wright et al.(\citealt{wright06}) 
this results in a luminosity distance of 102 Mpc
with a scale of 1 arcsec = 473 pc.\\

\section{Observations and data reduction}

\subsection{MUSE observations}

We observed NGC\,6240 on April 22, 2018, as part of the commissioning run of
the MUSE (Multi Unit Spectroscopic Explorer, Bacon et al. \citealt{bacon10},
\citealt{bacon14}) instrument in the Narrow-Field Mode (NFM)
with the four-laser adaptive optics system of ESO's Very Large Telescope
(VLT) unit telescope four (``Yepun'').
We used the peak of the southern emission region S1 (in the H band) 
as an on-axis tip-tilt object, centered by the large-scale pick-off. We exposed
four 500\,s science exposures on the center of NGC 6240 and another 500\,s
exposure on an offset sky field.
We rotated 90 deg between each on-target exposure. Spatial offsets of
about 1\farcs5 resulted in significantly larger coverage of
$\sim11\farcs5 \times 11\farcs5$ with some gaps at the field edges.
The NFM covers a field of
$7\farcs5\times7\farcs5$ on the sky, sampled at about 25 mas. 
A seeing of $0\farcs49 \pm 0\farcs07$ was measured with the
differential image motion monitor (DIMM) 
at the time of observations.
The target
was observed at an airmass of 1.13.
The wavelength range coverage is 4725 to 9350\,\AA{},
with a spectral resolution of about 2.5\,\AA{}.  
The spectra are sampled at 1.25\,\AA{} in dispersion direction 
and 0\farcs0253 in spatial direction.
 There is a gap in the spectrum between 5780 and
6050\,\AA{} because of the 
sodium laser-guide system.

\subsection{Data reduction} 
 
 We reduced the data using the MUSE pipeline development version 2.5.0
(Weilbacher et al.\citealt{weilbacher12}, \citealt{weilbacher14}) launched by
the EsoRex tool. We followed the usual steps of bias subtraction, flat-fielding
using lamp-flat, wavelength calibration, and twilight sky correction. We
also applied the usual illumination correction using a lamp-flat exposure taken
just minutes before the first science exposures. No standard star
was observed on April 22, so we used data of EG\,274
exposed under the same instrumental conditions in the night of April 20, 2018,
to compute atmospheric throughput
and characterize telluric absorption. We then created sky continuum and a
first-guess sky line flux table using the offset sky exposure. The NFM is
operated with an atmospheric dispersion compensator, so no software
correction was necessary. We flux-calibrated the data, subtracted the sky
continuum, and rescaled the sky line fluxes to the actual exposures and subtracted
them, before correcting the data for barycentric velocity (of 17.2 km/s). We
finally applied the astrometric distortion correction specific to NFM, computed
the relative offsets of the four exposures, and combined them into one final
datacube. This cube encompasses the data of all four exposures, but due to
the large offsets it has empty regions in the outer corners
(see Fig.~\ref{weilngc6240_map_Ha6745+GaiaDR2_v4_large.pdf}).
 
\subsection{Absolute positioning} 
 
  The absolute 
positions of the optical regions in NGC\,6240 were calibrated 
with respect to Gaia data.
For positioning, we shifted the pixel reference coordinates so that the
coordinates of the two sources from Gaia DR2
 (Lindegren et al.~\citealt{lindgren16}) in the field  
 overlap with peaks N and S2, as detected in the
$I$ band reconstructed from the MUSE data. Since the positions of the two
peaks can be made to agree with the Gaia DR2 positions, this implies
that our absolute astrometric accuracy is about 30\,mas (see Sect.
3.3).

 \section{Results}

 \subsection{Continuum and emission line morphology}
 
\begin{figure*}
\centering
\includegraphics[width=18. cm,angle=0]{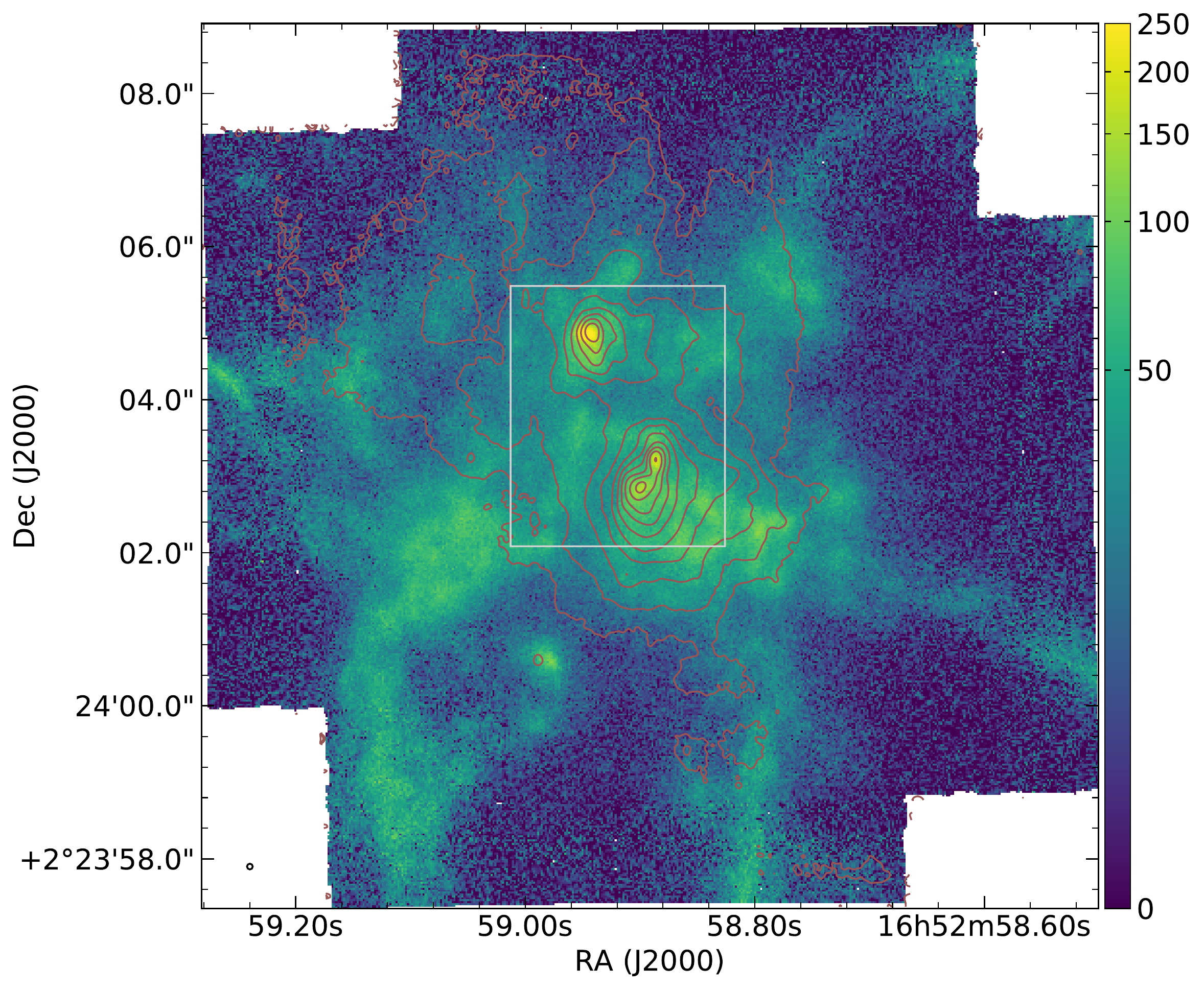}
\vspace*{-2mm} 
\caption{H$\alpha$/[NII] image 
%(at 6745 \AA{})
of the gas in NGC\,6240.
North is to the top and east to the left.
Overlaid are I band contour levels 
(see Fig.~\ref{weilngc6240_map_I+GaiaDR2_v4_ed.pdf}).
The white square indicates the size of the zoomed-in region in
Figs.~\ref{weilngc6240_map_I+GaiaDR2_v4_ed.pdf},
~\ref{weilngc6240_map_Ha6745+GaiaDR2_v4_ed.pdf},
~\ref{weilngc6240_map_V_SN003_clean+GaiaDR2_v4b_ed.pdf}, and
~\ref{weilngc6240_map_Sigma_SN003_clean+GaiaDR2_v4b_ed.pdf}.
The estimated spatial resolution at FWHM of 75 mas (35 pc)
is plotted in the bottom left corner of each image.
The flux is given in units of $10^{-20}$\,erg\,cm$^{-2}$\,s$^{-1}$\,pix$^{-1}$.}
\label{weilngc6240_map_Ha6745+GaiaDR2_v4_large.pdf}
\end{figure*}

\begin{figure*}
\centering
\includegraphics[width=11.5 cm,angle=0]{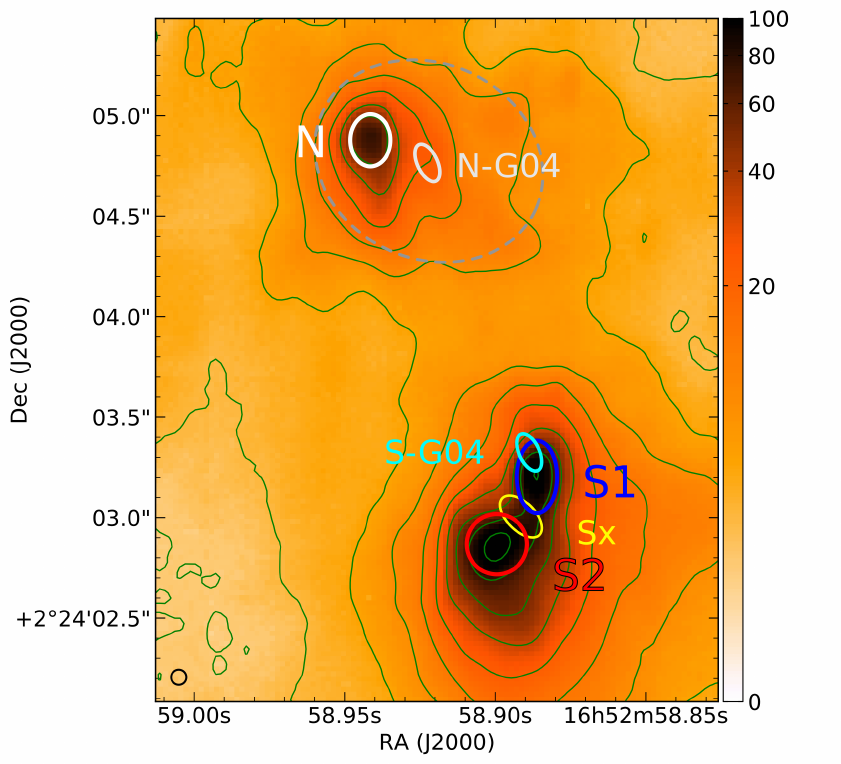}
\vspace*{-2mm} 
\caption{Cousins I-band image (intensities and
contours) of the zoomed-in region in NGC\,6240. The Cousins I band mostly traces 
the integrated distribution of the old stellar component. Spectra were extracted at the three locations corresponding to peaks of maximum intensity 
in the I-band map
(N, S1, S2, indicated by circles or ellipses).
Also indicated are the positions
of the two MERLIN and VLBA radio sources N-G04 and S-G04
(Gallimore et al.~\citealt{gallimore04}), and  the region Sx between the components S1 and S2.
The size of the northern stellar bulge component, based on the near-infrared CaII triplet lines
(see Sect. 3.5 and Figs.~\ref{weilngc6240_map_Ha6745+GaiaDR2_v4_ed.pdf},~\ref{weilngc6240_map_V_SN003_clean+GaiaDR2_v4b_ed.pdf},~\ref{weilngc6240_map_Sigma_SN003_clean+GaiaDR2_v4b_ed.pdf}),  is indicated by the dashed ellipse. 
The flux is given in units of $10^{-20}$\,erg\,cm$^{-2}$\,s$^{-1}$\,pix$^{-1}$.}
\label{weilngc6240_map_I+GaiaDR2_v4_ed.pdf}
\end{figure*}

\begin{figure*}
\centering
\includegraphics[width=11.5 cm,angle=0]{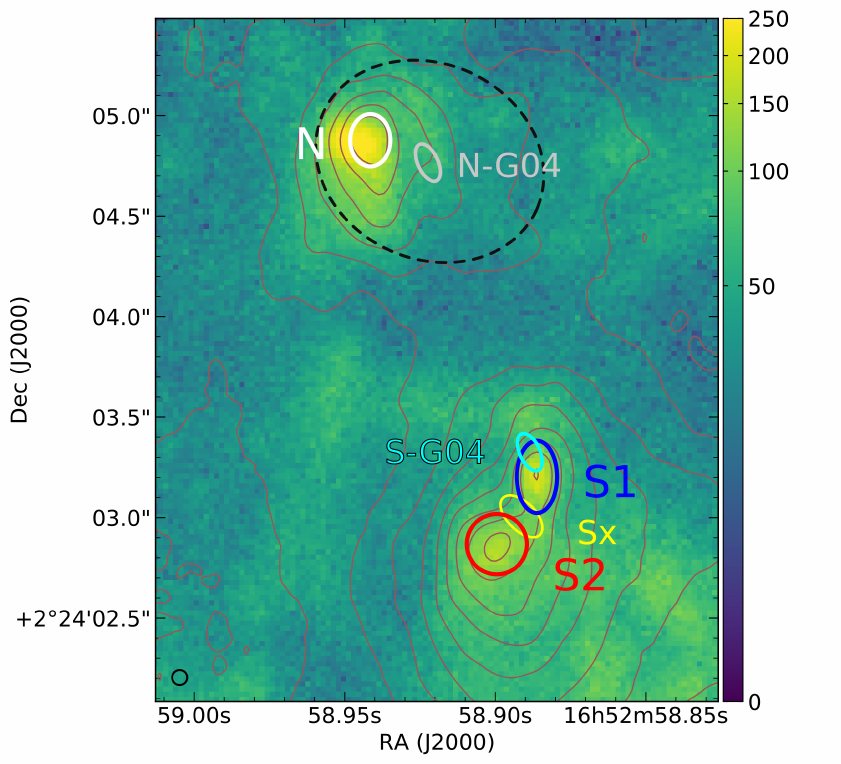}
\vspace*{-2mm} 
\caption{ 
H$\alpha$/[NII] image 
%(at 6745 \AA{})
of the  zoomed-in white square in
Fig.~\ref{weilngc6240_map_Ha6745+GaiaDR2_v4_large.pdf}.
Again, I-band contour levels are overlaid.
Spectra were extracted at the three locations corresponding to the peaks of maximum intensity 
in the I-band map
(N, S1, S2, indicated by circles or ellipses).
Also indicated are the positions
of the two MERLIN and VLBA radio sources N-G04 and S-G04
(Gallimore et al.~\citealt{gallimore04}). The flux is given in units of $10^{-20}$\,erg\,cm$^{-2}$\,s$^{-1}$\,pix$^{-1}$.}
\label{weilngc6240_map_Ha6745+GaiaDR2_v4_ed.pdf}
\end{figure*}

Figure~\ref{weilngc6240_map_Ha6745+GaiaDR2_v4_large.pdf}
shows the field of NGC\,6240
in the H$\alpha$/[NII] line complex
%at 6745 \AA{} 
covered by the four
overlapping MUSE observations. The H$\alpha$ line traces the distribution of the ionized gas.
Clearly, the emission line gas is distributed irregularly due to the merging process.
Overlaid are I-band contour levels.
The white square indicates the size of the zoomed-in area in 
Figs.~\ref{weilngc6240_map_I+GaiaDR2_v4_ed.pdf},
~\ref{weilngc6240_map_Ha6745+GaiaDR2_v4_ed.pdf},
~\ref{weilngc6240_map_V_SN003_clean+GaiaDR2_v4b_ed.pdf}, and
\ref{weilngc6240_map_Sigma_SN003_clean+GaiaDR2_v4b_ed.pdf}.


The inner region of NGC\,6240
is shown in Fig.~\ref{weilngc6240_map_I+GaiaDR2_v4_ed.pdf} in the Cousins I band
(effective central wavelength at $\lambda$8797\,\AA{}).
This wavelength range traces the old stellar component.
The circles and ellipses are based on the three maximum intensity positions in the
I band (N, S1, S2) where we extracted spectra.
In addition, we extracted a spectrum in the region Sx located between the emission regions S1 and S2.
Table~\ref{Coo-opt} gives the
coordinates of the I-band
continuum peaks, of Sx, and of the 
MERLIN and VLBA radio sources N-G04 and S-G04
where we extracted spectra as well.
\begin{table}
    \centering
       \leavevmode
       \tabcolsep1.5mm 
\caption{I-band continuum peaks of the northern component N 
and the southern components S1 and S2.
Also listed is the position Sx where we took
a spectrum between the components S1 and S2.
The optical errors are 1$\sigma$ errors (1D modeling).
The
MERLIN and VLBA radio sources N-G04 and S-G04
at 2.4 and 1.7\,GHZ are also given (Gallimore et al.\citealt{gallimore04}).}
\begin{tabular}{lcc}
 \htopline
\multicolumn{1}{l}{ID} & RA\,(J2000) & DEC\,(J2000) \\
\hmidline   
\noalign{\smallskip}
N   & 16:52:58.943$\pm{}$.0005 & +2:24:04.89$\pm{}$.003 \\
N-G04  & 16:52:58.924$\pm{}$.0003 & +2:24:04.766$\pm{}$.002\\
S-G04  & 16:52:58.890$\pm{}$.0003 & +2:24:03.337$\pm{}$.002\\
S1       & 16:52:58.888$\pm{}$.0005 & +2:24:03.22$\pm{}$.003 \\
Sx      & 16:52:58.893$\pm{}$.0015 & +2:24:03.02$\pm{}$.01 \\
S2  & 16:52:58.901$\pm{}$.0015 & +2:24:02.88$\pm{}$.01 \\
\noalign{\smallskip}
\hbotline  
\end{tabular}
\label{Coo-opt}
\end{table}

The zoomed-in region of NGC\,6240 in the H$\alpha$/[NII] line 
is shown in Fig.~\ref{weilngc6240_map_Ha6745+GaiaDR2_v4_ed.pdf}.
The maxima of the southern peaks in the I band correspond to the peaks of H$\alpha$ emission. However, there is a slight
offset for the northern peak.
The projected distance between the optical maxima of the northern N
and S1 component (I band) amounts to 1.81$\pm$0.03\,arcsec 
(856$\pm$14\,pc); the projected distance
between the two southern components S1 and S2 amounts to 0.42 $\pm$0.03\,arcsec
(198$\pm$14\,pc).

\subsection{Spatial resolution}
 
\begin{figure*}
\centering
\includegraphics[width=13.cm,angle=0]{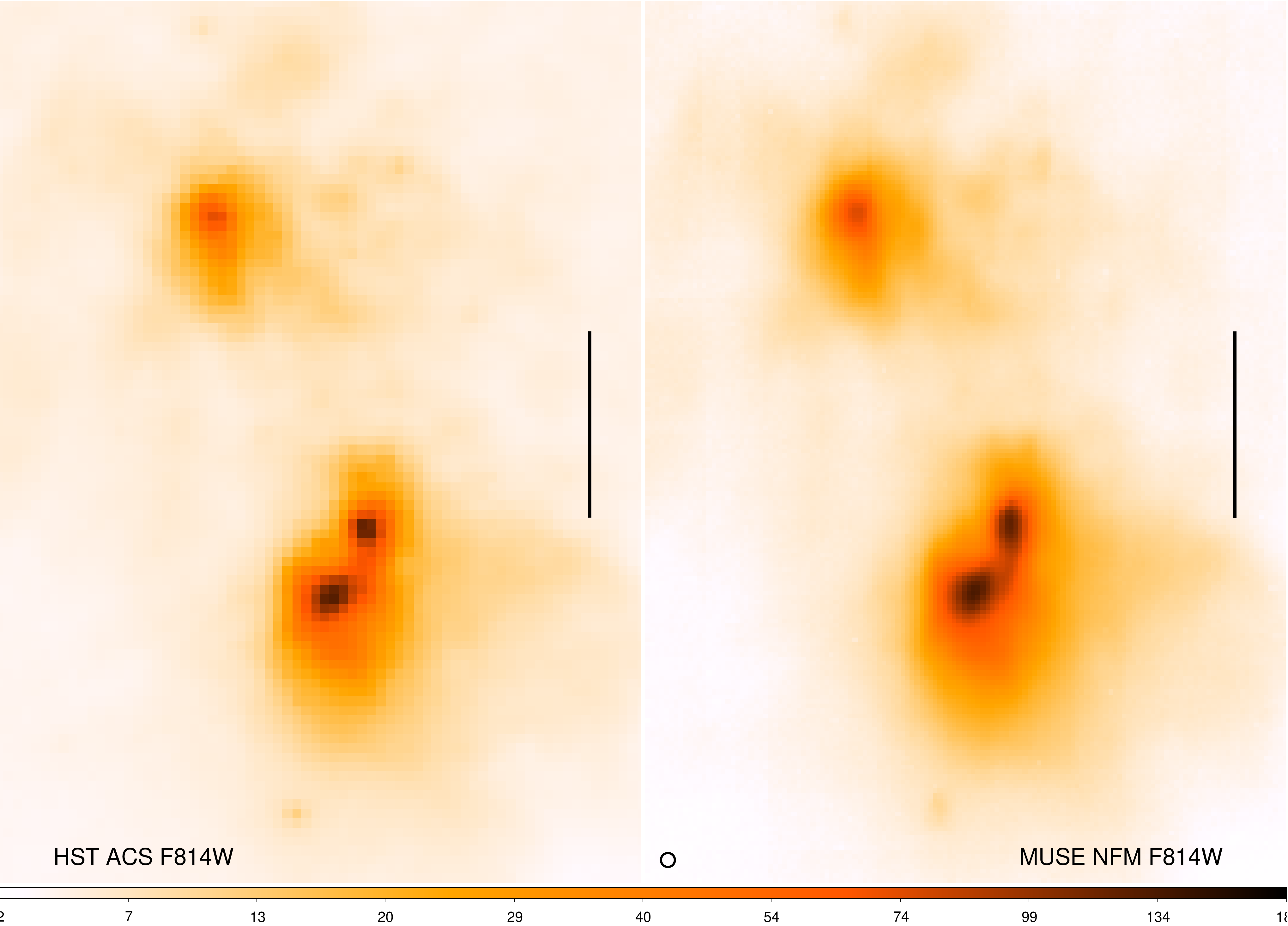}
\vspace*{3mm} 
\caption{Images of NGC\,6240 taken with the HST-ACS and F814W filter and with MUSE
in the NFM for the I-band. The black line is 1 arcsec long.
An inverse asinh scaling was used for the color map.}
\label{weilcomparison_HST-ACS+MUSE-NFM_F814W_no_overplots_v4.pdf}
\end{figure*}

The characterization of the point spread function (PSF) of the NFM data is still
ongoing. 
Our field does not contain any unresolved objects and we do not know
of a higher resolution image of the field, so we cannot derive a firm
measurement of the FWHM of our data.
However, we reconstructed an image corresponding to the HST ACS F814W filter 
from the MUSE data, and we can visually see smaller details than in the HST
ACS data (FWHM of 0\farcs1, HST proposal ID 10595, PI A.\ Evans) taken in that
filter. We therefore assume an approximate core resolution of $\sim75$ mas
in the MUSE NFM data at the red end of the wavelength range. 
We present observations
of the inner region of NGC\,6240 taken with the ACS camera on board the HST and for comparison
a MUSE image in Fig.~\ref{weilcomparison_HST-ACS+MUSE-NFM_F814W_no_overplots_v4.pdf}.
At the distance of
NGC 6240 we therefore cover a region of 3.5\,kpc in each exposure with a
spatial resolution of about 35\,pc.

 \subsection{Absolute positional accuracy}

\begin{figure*}
\centering
\includegraphics[width=13.cm,angle=0]{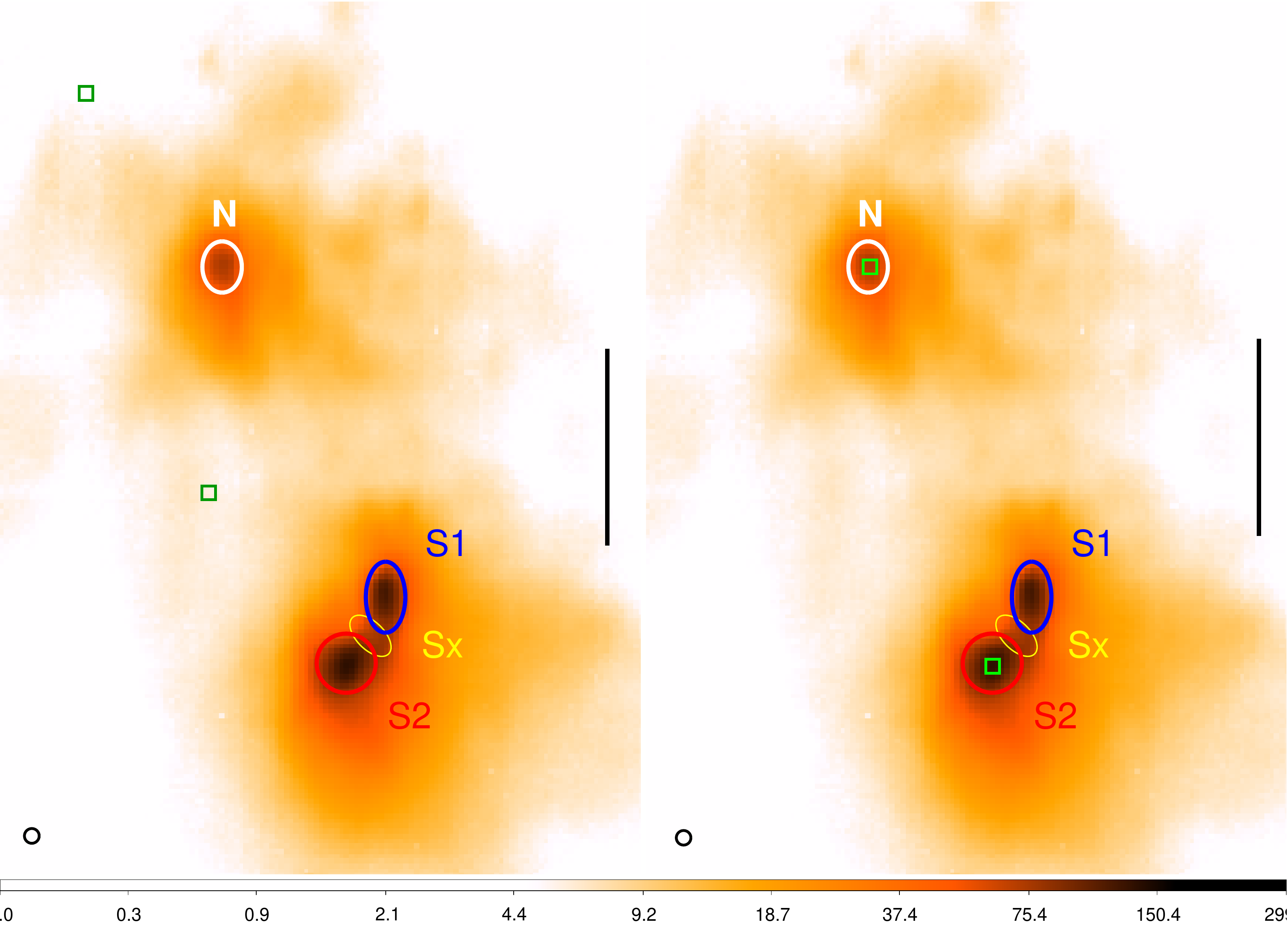}
\vspace*{3mm} 
\caption{Positions of Gaia sources (green squares) and  MUSE sources before (left) and
after (right) absolute calibration  of the MUSE data. The black line is 1 arcsec long.
An inverse log scaling was used for the color map.}
\label{weilGaia_positions_v4.pdf}
\end{figure*}

The absolute positional accuracy of our optical positions in NGC\,6240 is very high as they 
have been calibrated with respect to Gaia data.
Therefore, the given positions of the emission and the absorption regions 
have an absolute error of only 30\,mas. 
We show the Gaia positions of the regions N and S2 
before and after absolute calibration of our MUSE data in Fig.~\ref{weilGaia_positions_v4.pdf}.
The optical continuum and emission line peak S1 coincides with 
the center of the blueshifted stellar component
(Fig.~\ref{weilngc6240_map_V_SN003_clean+GaiaDR2_v4b_ed.pdf}) and with
the maximum of the stellar velocity dispersion
 (Fig.~\ref{weilngc6240_map_Sigma_SN003_clean+GaiaDR2_v4b_ed.pdf}), and
with the southern radio component S-G04 based on the MERLIN and VLBA data
(Gallimore et al.~\citealt{gallimore04}). All these
positions agree within   0.08 arcsec.

\subsection{Emission line spectroscopy}

\begin{figure*}
\centering
\includegraphics[width=17. cm,angle=-90]{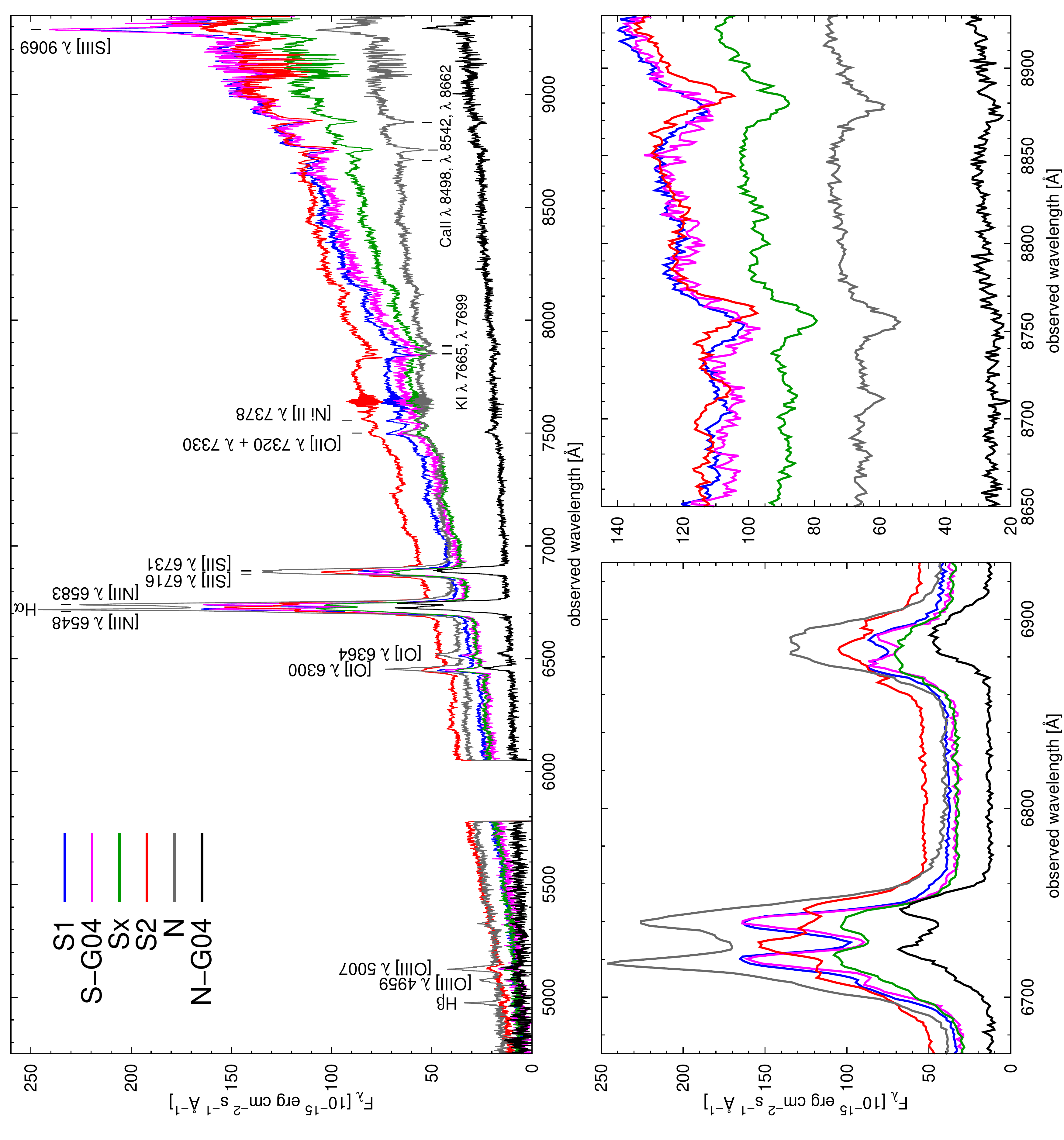}
\caption{Spectra of the northern N (gray) and southern S1 (blue) and
S2 (red) emission regions of the northern N-G04 (black) and southern S-G04 (pink)
MERLIN and VLBA radio positions, and the region Sx between the southern components (green).
In the lower panels are shown enlargements of the H$\alpha$/[SII] complex and of the 
near-infrared CaII triplet lines.
}
\label{ochmall_multi.pdf}
\end{figure*}

We present the spectra of the 
emission regions N, S1, S2, and of the region Sx between S1  and S2 
in Fig.~\ref{ochmall_multi.pdf}.
In addition, we show the spectra extracted at the positions of the northern
(N-G04) and southern (S-G04) MERLIN and VLBA radio positions (Gallimore et al.~\citealt{gallimore04}).
The corresponding spatial regions (ellipses) we used to extract the
spectral flux of the individual regions
are shown in
Figs.~\ref{weilngc6240_map_I+GaiaDR2_v4_ed.pdf} and
\ref{weilngc6240_map_Ha6745+GaiaDR2_v4_ed.pdf}.
The sizes and orientations of the radio ellipses  adopted are from the MERLIN observations
(Gallimore et al.~\citealt{gallimore04}). The centers of the N, S1, and S2 ellipses are based on
the maxima of the I-band contour levels. The diameters of the ellipses correspond to a value of
three to five times that of our spatial resolution. The Sx ellipse was chosen in order to fit between those of S1 and S2.
The exact sizes of the ellipses of the individual line emitting regions are 
given in Table~\ref{ell_sizes}. 
\begin{table}%[htbp]
    \centering
       \leavevmode
       \tabcolsep1.5mm 
\caption{
Sizes of the ellipses of the individual line emitting regions.}
\begin{tabular}{lccc}
 \htopline
\multicolumn{1}{l}{component} & \mcc{minor axis}  &  \mcc{major axis} & \mcc{angle}\\
\hspace{3mm}        & \mcc{[acrsec]} & \mcc{[arcsec]} & \mcc{[deg]} \\
\hmidline   
\noalign{\smallskip}
N            & 0.10 & 0.13 & 0 \\
N-G04    & 0.05 & 0.10 & 25 \\
S-G04    & 0.05 & 0.10 & 25 \\
S1  & 0.10 & 0.18 & 0\\
Sx           & 0.07 & 0.13 & 45 \\
S2           & 0.15 & 0.15 & 0\\
\noalign{\smallskip}
\hbotline  
\end{tabular}
\label{ell_sizes}
\end{table}
All the observed spectra show a strong red continuum because of heavy dust absorption.

\begin{figure*}
\centering
\includegraphics[height=14.cm,angle=-90]{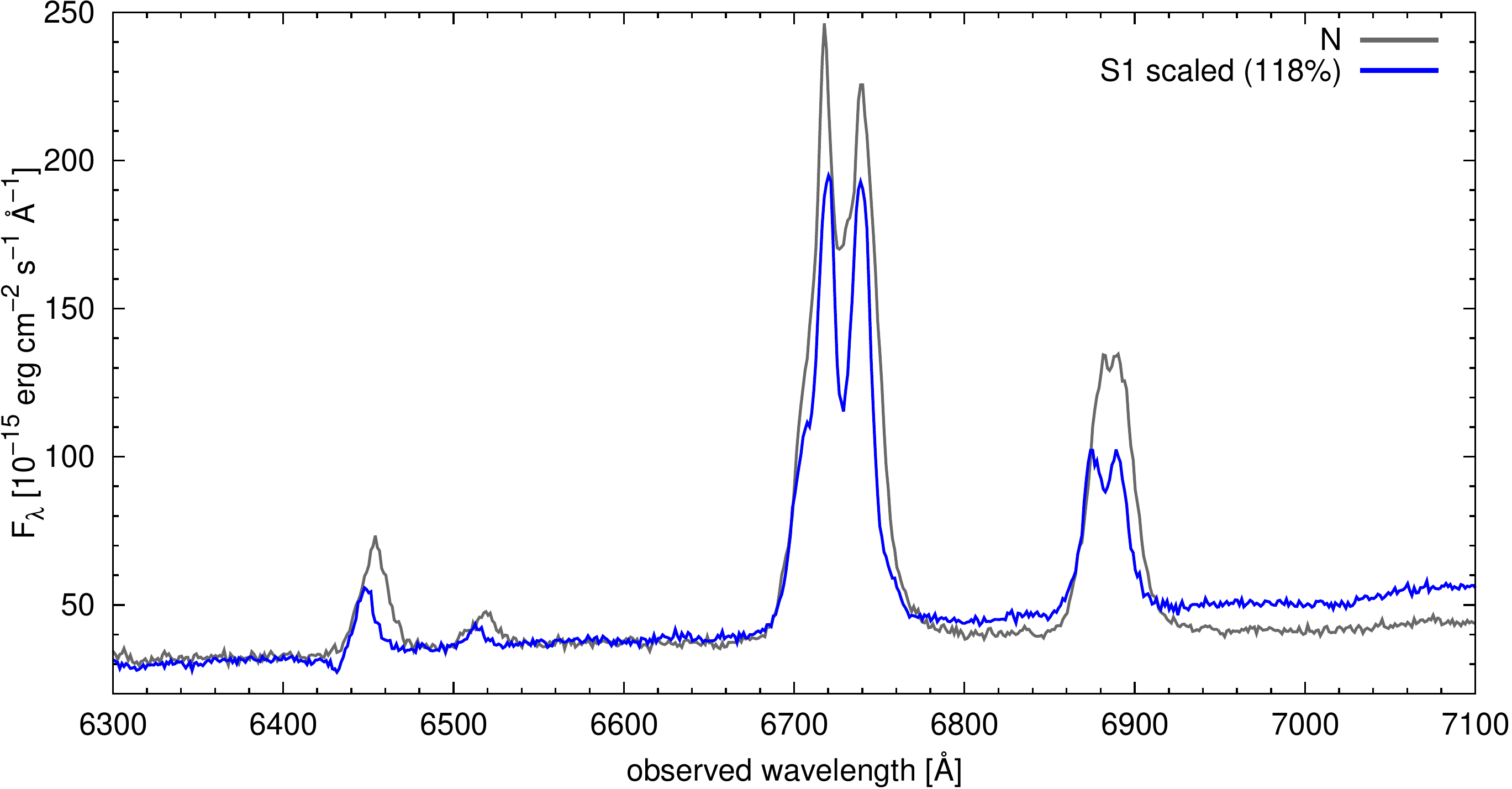}
\caption{Spectra of the emission regions N (gray) and S1 (blue)
scaled to the same H$\alpha$ velocity. 
}
\label{ochmN_S1.pdf}
\end{figure*}

We determined the redshifts, the line intensities,
and the line widths (FWHM) of the strongest emission lines
in the individual line emitting regions (see Appendix A).
We fitted the H$\alpha$ 
line complex by multiple Gaussians with the IRAF task SPLOT.
Here we set the additional condition that both [NII] lines must hold
the same width.
The redshifts of the individual emission regions
were fixed on the basis of the 
H$\alpha$ line, the strongest emission line.
The Balmer lines as well as the [NII] and [OIII]5007 lines
show very similar redshifts at the individual positions in NGC\,6240.
However, the [OI]6300 line deviates by up to 150 km s$^{-1}$ to higher
and to lower redshifts, indicating that this line originates at different layers 
in NGC\,6240 compared to the other emission lines.
The [OI]/H$\alpha$/[SII] complex of the components N and S1 is shown in
Fig.~\ref{ochmN_S1.pdf}. 
The spectra have been shifted to the same H$\alpha$ velocity and scaled to similar
H$\alpha$ intensities. It can be  seen that the relative velocities of the
[OI] and [SII] lines do not agree with those of the  H$\alpha$ and
[NII] lines.

\begin{figure*}
\centering
\includegraphics[height=14.cm,angle=-90]{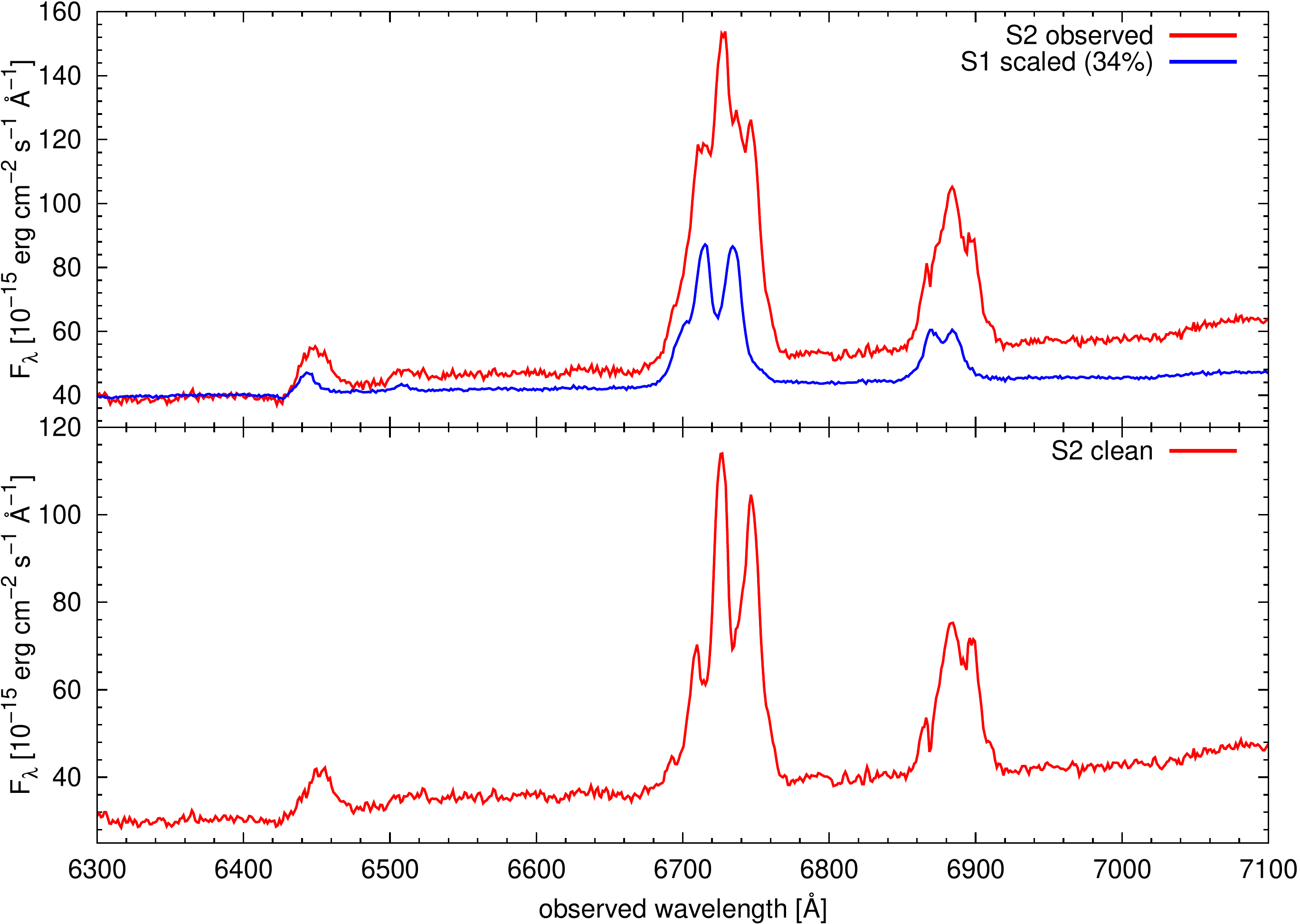}
\caption{
Raw spectrum of S2 (red) and a scaled (by 34 \%) S1
spectrum shifted by 222 km s$^{-1}$ (blue).
The difference spectrum at the bottom
shows the clean S2 spectrum.
}
\label{ochmdiff_S2_S1.pdf}
\end{figure*}

It is evident 
from the H$\alpha$/[NII] and [SII] line spectra (Fig.~\ref{ochmall_multi.pdf})
that the observed emission line profiles of the regions S2 and Sx are quite
complex. More precisely, each spectrum is a superposition of at least
two emission components (Figs.~\ref{ochmall_multi.pdf} and ~\ref{ochmdiff_S2_S1.pdf}).
On the other hand, the H$\alpha$/[NII] complex of the S1 region 
can easily be fitted with only one emission component. To decompose the S2 and Sx spectra, we made the
assumption that one of the components shows emission line profiles with
a similar spectral shape to that of the S1 complex. Thus, we subtracted a series of
shifted and scaled S1 spectra from the S2 and Sx spectra until the 
difference spectrum resulted in an  H$\alpha$/[NII] complex where all the lines showed
similar line widths and where the [NII] line ratio corresponded to the theoretical ratio of 1/3. 
As our best solution we subtracted 
a scaled (by 34 \%)
and shifted (by 222 km s$^{-1}$) S1 spectrum from the H$\alpha$/[NII] and [SII] complex in S2
(see Fig.~\ref{ochmdiff_S2_S1.pdf}) and  in Sx.
The resulting clean S2 spectrum is a better match
for a single nucleus spectrum.
From here on the spectra of S2 and  Sx refer to this clean version
(Tables~\ref{BPT_tab}, ~\ref{abs_vel}, ~\ref{Sbetween_emlines}, ~\ref{S2_emlines}).

We derived the dust extinction in the individual emitting regions
based on the Balmer decrement H$\alpha$/H$\beta${}
using the formulas given by
Dominguez et al.\cite{dominguez13}. 
We calculated high extinction values A$_{V}$   (on the order
of ten) at all five component positions except at the northern region N, indicating
a lower extinction of only A$_{V}$=5 (see Table~\ref{BPT_tab}).
Furthermore, the dust extinction could in principle also be estimated from the spectral energy
distribution.
We see some trends between the two estimates regarding the amount of dust
in NGC\,6240. The northern 
component N shows the smallest Balmer decrement
and the flattest spectral continuum flux distribution, while
the component S1 shows the highest Balmer decrement 
and a very steep continuum slope.
However, it is known that
the stellar continuum reddening shows no clear correlation with the
dust content, suggesting that the distribution of stellar reddening does not act as a
good tracer of the overall dust content (e.g., Kreckel et al.
\citealt{kreckel13}).

We determined the ionization level of the individual emission regions 
based on optical emission-line ratios (i.e., diagnostic diagrams based on the
line intensity ratios of [OIII]5007/H$\beta$ versus
[NII]6584/H$\alpha$;
Kauffmann et al.\citealt{kauffmann03}, Kewley
et al.\citealt{kewley06}).
The line ratios of the individual emission regions and of the radio sources
are given in Table~\ref{BPT_tab}.
\begin{table*}
%\begin{table*}%[htbp]
    \centering
       \leavevmode
       \tabcolsep1.5mm 
\caption{
Balmer decrement and line intensity ratios of  \Nl{[N}{ii]}{6584}/H$\alpha$ and
\Nl{[O}{iii]}{5007}/H$\beta$ (BPT diagram) for the components N, S1, S2, Sx,
and  for the spectra taken at the radio positions.}
\begin{tabular}{lcccccc}
 \htopline
\multicolumn{1}{l}{Intensity ratio} & N & N-G04 & S-G04 & S1 &  Sx & S2 \\
\hmidline   
\noalign{\smallskip}
H$\alpha$/H$\beta$       & 12.1   & 54.:  & 73.:  &  64. &  27. & 45.: \\
A$_{V}$                  &  5.0   &  10.2:& 11.2: &  10.8 & 7.8 &  9.5: \\
log([NII]/H$\alpha$)     & 0.327$\pm{}$0.018  & 0.064$\pm{}$0.018 & 0.055$\pm{}$0.018 & 0.047$\pm{}$0.019 & 0.031$\pm{}$0.020  &  0.068$\pm{}$0.019 \\
log([OIII]/H$\beta$)     & 0.371$\pm{}$0.025  & 0.710$\pm{}$0.146 & 0.577$\pm{}$0.071 & 0.507$\pm{}$0.025 & 0.403$\pm{}$0.220  &  0.434$\pm{}$0.066 \\  
\noalign{\smallskip}
\hbotline  
\end{tabular}
\label{BPT_tab}
\end{table*}
The line ratios of all investigated spectra (Fig.~\ref{ochmall_multi.pdf})
correspond to LINER-like objects.
Although all spectra fall in the LINER region in the diagnostic diagram,
 it is unlikely that all of them are 
caused by  photoionization from individual
active nuclei (see Sect. 4.1). 
Furthermore, we observe emission line widths of 500 to 700 km s$^{-1}$ (Appendix A).
Such
high line widths are rarely
observed in LINER nuclei.

\subsection{Stellar kinematics}

To derive stellar kinematics around the continuum peaks, we processed the MUSE
cube using pPXFv6.7.12 (Cappellari \& Emsellem\citealt{cappellari04},
Cappellari\citealt{cappellari17}).
Since the S/N was only sufficient in the red part to see
continuum features, we restricted the wavelength range to 8350-9200\,\AA{}.
This wavelength range includes the near-infrared Ca II triplet lines
 (8498, 8542, 8662\,\AA{}).
We used a set of 53 high-quality stars from the Indo-US
library as stellar templates (Valdes et al.\citealt{valdes04},
Shetty \& Cappellari\citealt{shetty15},
and Gu\'erou et al.\citealt{guerou17}). This library
provides a high enough resolution compared to the MUSE data, and fully covers the
wavelength range of interest. 
We convolved the templates to
an estimate of the average instrumental width and then used pPXF to fit velocity
and velocity dispersion for every spectrum. Since the fit only converges to
sensible values of continuum kinematics around the peaks, we cleaned up strongly
deviant fits and those where the outputs are identical to the starting values.

\begin{figure*}
\centering
\includegraphics[width=12.cm,angle=0]{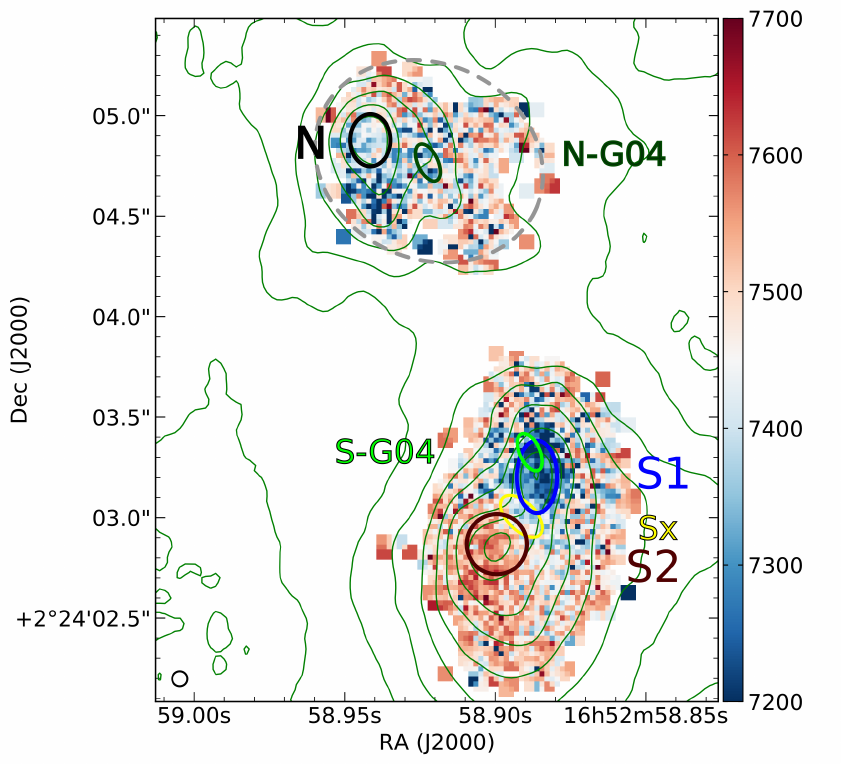} 
\vspace*{-2mm} 
\caption{Redshift of the stellar component based on the CaII IR triplet lines
 in units of km s$^{-1}$
 for the northern and southern nuclear bulge regions.
White regions represent locations where the S/N
was not high enough to obtain good fits.
The size of the northern stellar bulge component is indicated by the dashed ellipse 
(see also Fig.~\ref{weilngc6240_map_Sigma_SN003_clean+GaiaDR2_v4b_ed.pdf}).
}
\label{weilngc6240_map_V_SN003_clean+GaiaDR2_v4b_ed.pdf}
\end{figure*}

\begin{figure*}
\centering
\includegraphics[width=12.cm,angle=0]{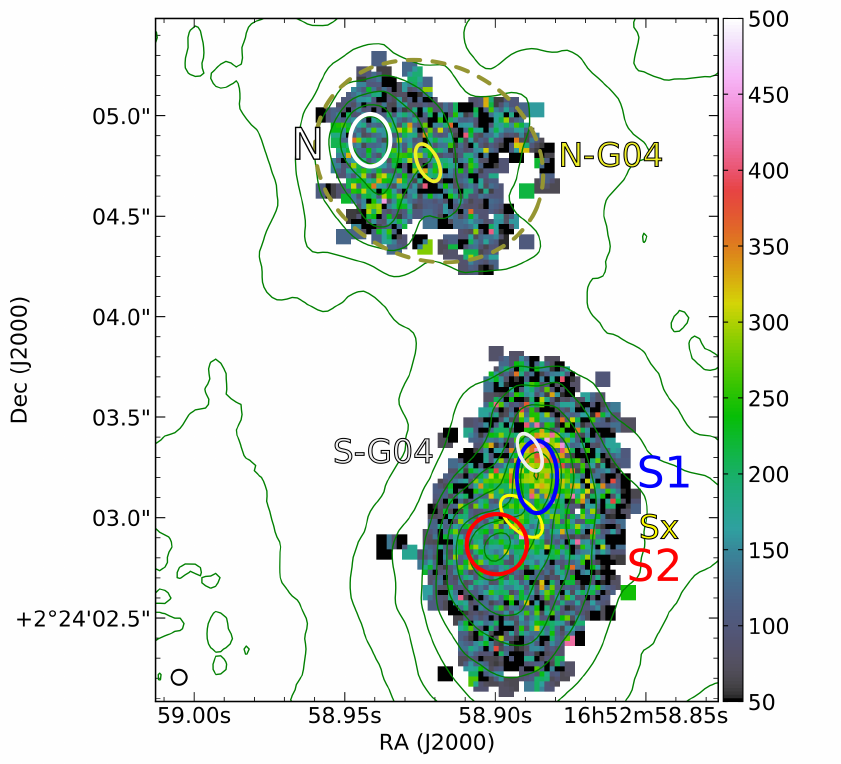} 
\vspace*{-2mm} 
\caption{Stellar dispersion map 
of the northern and southern nuclear bulge regions
 based on the CaII IR triplet lines
 in units of km s$^{-1}$.
White regions represent locations where the S/N
was not high enough to obtain good fits.
}
\label{weilngc6240_map_Sigma_SN003_clean+GaiaDR2_v4b_ed.pdf}
\end{figure*}

Based on the presence of the
near-infrared  calcium II triplet lines, which are very strong in K giant stars and can 
be observed in high-extinction regions such as galactic bulges, we deduced the existence 
of old stellar bulge components in NGC\,6240 (e.g., Vasquez et al.\citealt{vasquez15} 
and references therein). Consequently, we used the calcium II triplet lines to derive 
the redshift and the stellar dispersion of the old stellar bulge components.
The resulting maps are shown in
 Figs.~\ref{weilngc6240_map_V_SN003_clean+GaiaDR2_v4b_ed.pdf}
and \ref{weilngc6240_map_Sigma_SN003_clean+GaiaDR2_v4b_ed.pdf}.
A strong signal of an old stellar bulge component is seen in the northern region next
to the northern emission region N, and at each of the two southern emission line regions S1 and S2.
We present in Table~\ref{abs_vel} the absorption line velocities and velocity dispersions based on the CaII
IR triplet lines 
for the components N, S1, and S2, and for the positions of the
MERLIN and VLBA radio sources N-G04 and S-G04 and at Sx.
The region S2 shows a small gradient in the absorption velocity and in the stellar dispersion.
(Figs.~\ref{weilngc6240_map_V_SN003_clean+GaiaDR2_v4b_ed.pdf} and \ref{weilngc6240_map_Sigma_SN003_clean+GaiaDR2_v4b_ed.pdf}).
Therefore, we measured these S2 component values using a radius that is 20 \% smaller in order to exclude the gradient.
For comparison we present the emission line velocities and the 
line widths (FWHM) based on the H$\alpha$ emission line for the same regions (see Table~\ref{abs_vel}).
\begin{table*}%[htbp]
    \centering
       \leavevmode
       \tabcolsep1.5mm 
\caption{
Absorption line velocities and velocity dispersions based on the CaII
IR triplet lines 
for the components N, S1, and S2, and on the positions of the
MERLIN and VLBA radio sources N-G04 and S-G04.
For comparison we present the emission line velocities and the 
line widths (FWHM and $\sigma$) based on the H$\alpha$ line.}
\begin{tabular}{lccccc}
 \htopline
\multicolumn{1}{l}{component} & \mcc{v\,(CaII)}  &  \mcc{$\sigma$\,(CaII)} & \mcc{v\,(H$\alpha$)} & \mcc{FWHM\,(H$\alpha$)} & \mcc{$\sigma$\,(H$\alpha$)} \\
\hspace{3mm}        & \mcc{[\kms{}]} & \mcc{[\kms{}]} & \mcc{[\kms{}]} & \mcc{[\kms{}]} & \mcc{[\kms{}]} \\
\hmidline   
\noalign{\smallskip}
N             & 7404.$\pm{}$13. & 157.$\pm{}$13. & 7117.4$\pm{}$5. &  602.$\pm{}$16. &  256.$\pm{}$8.\\
N-G04   & 7297.$\pm{}$41. & 282.$\pm{}$31. &  7473.9$\pm{}$5. &  677.$\pm{}$20.   &  288.$\pm{}$9.\\
S-G04     & 7202.$\pm{}$21. & 316..$\pm{}$35. &  7176.8$\pm{}$9. & 578.$\pm{}$17.    &  246.$\pm{}$8.\\
S1           & 7260.$\pm{}$29. & 297.$\pm{}$29. & 7163.1$\pm{}$5  &  605.$\pm{}$22.&  257.$\pm{}$10.\\
Sx           & 7422.$\pm{}$20. & 203.$\pm{}$28. & 7528.8$\pm{}$9. &  553.$\pm{}$22. &  235.$\pm{}$10.\\
S2          & 7530.$\pm{}$15. & 180.$\pm{}$30. & 7473.9$\pm{}$9. &  508.$\pm{}$15.&  216.$\pm{}$8.\\
\noalign{\smallskip}
\hbotline  
\end{tabular}
\label{abs_vel}
\end{table*}

\subsection{Black hole masses}
We determined stellar velocity dispersions at the positions of the optical emission maxima
S1 and S2
and at the radio positions (Fig.~\ref{weilngc6240_map_Sigma_SN003_clean+GaiaDR2_v4b_ed.pdf},
Table~\ref{abs_vel}). Hereafter, we assume that the components S-G04 and S1  originate from the same 
nucleus as their positions overlap. 
We identify each of the emission regions N-G04, S-G04/S1, and S2 to be separate nuclei in the centers 
of independent galactic bulges, hence each of these emission regions is associated with a SMBH.
Based on the radio emission, this assumption is obviously justified for N-G04 and S-G04/S1.
Although we are not able to resolve the sphere of influence for a black hole in S2 and do not observe 
an accretion signature, we find strong evidence for the existence of an independent nucleus S2 
(see Sects. 4.2, 4.3, and 4.4).
We calculate
corresponding black hole masses $M_\text{BH}$ for each of the nuclei by using the tight correlation 
between the stellar velocity dispersion $\sigma_\text{star}$ 
in galaxy bulges and $M_\text{BH}$
in inactive and active galaxies
(e.g., Gebhardt et al.\citealt{gebhardt00} and Greene \& Ho \citealt{greene06}).
The observed velocity dispersion values correspond to black hole masses of
\vspace*{0.1cm} \\
$\text{M(N-G04)}\,\,\,\,\,\,\,=3.6\pm0.8\times10^{8} M_{\odot}$,\\
$\text{M(S-G04/S1)}=7.1\pm 0.8\times10^{8} M_{\odot}$,\\
$\text{M(S2)}\,\,\,\,\,\,\,\,\,\,\,\,\,\,\,\,\,=9.0\pm0.7\times10^{7} M_{\odot}$
\vspace*{0.1cm} \\
when using the $M_\text{BH}$--$\sigma_\text{star}$ diagram 
of Greene \& Ho (\citealt{greene06}). The resulting 
black hole mass ratio of the triple system is \vspace*{0.1cm}\\
{}[4 : 8 : 1] for [N-G04 : S-G04/S1 : S2].\vspace*{0.1cm}\\
We did not derive a black hole mass for component N as it is not connected with a 
separate nucleus, but is located in the foreground at the edge of the northern bulge (see Sect. 4.4).\\
Deriving the orbital mass $M_\text{orb}$ (Petrosyan\citealt{petrosyan83}) offers another way
to estimate the integrated mass of a double nucleus system.
Making the assumption that both nuclei S1 and S2 are in circular motion about a common
center and knowing the radial velocity difference between the nuclei (270 km s$^{-1}$) and 
their projected distance (198 pc), we 
calculated the orbital mass $M_\text{orb}$ of the S1-S2 system
using the equation (Petrosyan\citealt{petrosyan83}) \\
\begin{align}
M_\text{orb}=\frac{32}{3\pi}(\Delta V)^2_\text{pr}\frac{D_\text{pr}}{G}  
\label{eq:orbital_mass},
\end{align}
 where $\frac{32}{3\pi}$ is a mean projection factor.
We derive an orbital mass $M_\text{orb}$ of $1.2\times 10^{10} M_{\odot}$, 
Since the orbital mass is usually higher
than the mass of the two individual nuclei, this result fits with the black hole masses
derived on the basis of the stellar velocity dispersion.

\section{Discussion}

\subsection{Excitation of the emission lines in the central region of NGC\,6240}

\begin{figure}
\centering
\includegraphics[width=8.5cm,angle=-90]{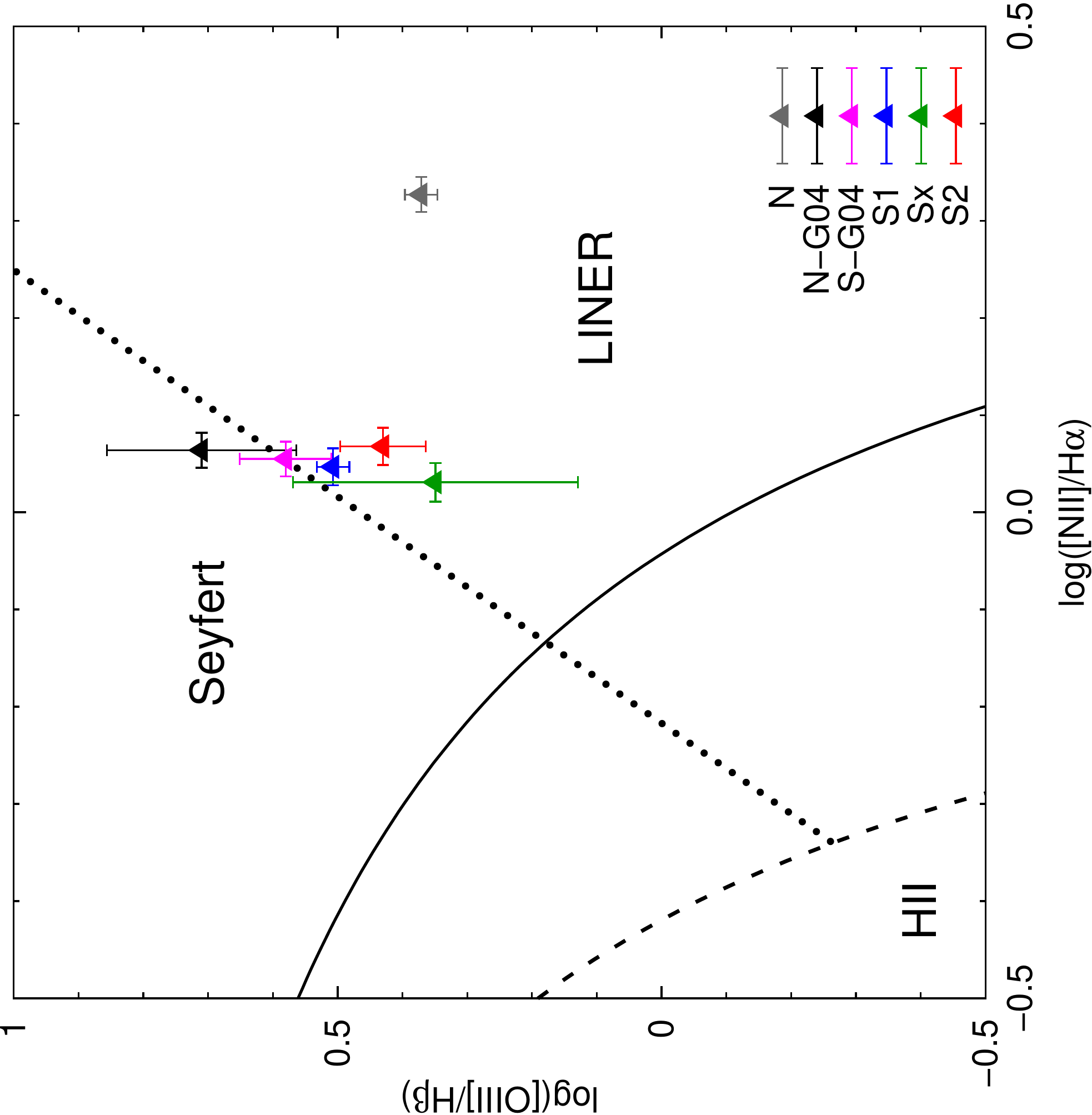}
\caption{Diagnostic BPT diagram of [OIII]{5007}/H$\beta$ vs.  [NII]{6584}/H$\alpha$ for individual 
emission line regions in NGC\,6240.
The dashed lines and solid lines are the dividing lines between
AGN (upper right) and HII-region galaxies (lower left) 
(Kauffmann et al.\citealt{kauffmann03}, Kewley et al.\citealt{kewley06}). 
The dotted line is the dividing line between Seyfert and LINER nuclei.
}
\label{ochmbpt.pdf}
\end{figure}

NGC\,6240 is classified
as a LINER type in a previous work (Veilleux et al.~\citealt{veilleux95})
based on an integrated optical spectrum of the central region.
Instead, we determined the activity level of all the individual central emission regions 
based on their optical emission-line  ratios (Table~\ref{BPT_tab}).
The line ratios of all six investigated spectra (Fig.~\ref{ochmall_multi.pdf})
correspond to LINER-like objects. 
Figure~\ref{ochmbpt.pdf} shows the 
[OIII]{5007}/H$\beta$ versus [NII]{6584}/H$\alpha$ 
line diagnostic diagram 
for the individual
emission regions. 
However, all the observed line ratios are consistent
with shock heating models as well
(see the models of Allen et al.\citealt{allen08}).

It is unlikely that all the individual spectra in NGC\,6240 are 
caused by  photoionization from six individual active nuclei. 
 Alternatively, the observed line ratios can instead be explained by shock heating
(e.g., Dopita \& Sutherland\citealt{dopita95},
Monreal-Ibero et al.\citealt{monreal10},
 Marziani et al.\citealt{marziani17}).
Narrow emission line widths (FWHM) of 500 to 700 km s$^{-1}$, like those seen in NGC\,6240,
are rarely observed in Seyfert and LINER nuclei (e.g., Ho et al.\citealt{ho03},
 Zhang et al.\citealt{zhang13}). The nuclei S1 and S2 exhibit line widths of 600 km s$^{-1}$;
the Balmer lines of the northern regions indicate emission line widths of
the same order.

In the northern region N the line widths of the nitrogen and oxygen
lines are even broader (by 400 km s$^{-1}$) in comparison to H$\alpha$ (see Table~\ref{N_emlines}).
These observed line profiles might be a superposition of many components
originating in different regions.
Emission lines broader than 600 km s$^{-1}$ are known from radio galaxies and
luminous infrared galaxies (LIRGs), due to additional motions caused by radio
jets and/or galaxy merging processes including outflowing processes
(e.g., Buchanan et al.,\citealt{buchanan06}).
It is therefore intriguing that a small-scale radio jet has been detected at the position of
N-G04 (Gallimore et al.~\citealt{gallimore04}, Hagiwara et al.~\citealt{hagiwara11}).
NGC\,6240 is both a merger and an LIRG and as such displays
broad line widths caused by a superposition of different components
including turbulent motions.
Radio supernova and narrow $H_{2}O$ maser lines near the S-G04 region suggest some
circumnuclear star formation and dense molecular gas (Hagiwara et al.~\citealt{hagiwara11}).
Hence, all observed emission line spectra in NGC\,6240 are almost
certainly caused by shock heating. 
An independent confirmation of shock heated gas comes from spatially resolved 
hard X-ray emission in the central 5 kpc detected on deep Chandra images
(Wang et al.~\citealt{wang14}).

\subsection{Black hole masses}

We derived the stellar velocity dispersions at the
northern and southern radio sources and at S2. Afterwards we
determined black hole masses of $\text{M(N-G04)}=3.6\pm0.8\times10^{8} M_{\odot}$ ,
$\text{M(S-G04/S1)}=7.1\pm 0.8\times10^{8} M_{\odot}$,
and $\text{M(S2)}=9.0\pm0.7\times10^{7} M_{\odot}$
based on the $M_\text{BH}$--$\sigma_\text{star}$ diagram of Greene \& Ho (\citealt{greene06}).  
Furthermore,  we calculated the orbital mass $M_\text{orb}$ of the S1--S2 system (using Eq. \eqref{eq:orbital_mass})
to be $M_\text{orb} = 1.2\times 10^{10} M_{\odot}$.
Assuming a relaxed gravitationally bound S1--S2 system, and considering an S2 radius of
$\sim$100 pc and velocity dispersion of
180\,km s$^{-1}$, an isothermal sphere would have a mass of $\sim 7 \cdot 10^{8}$\,M$_\odot$.
The size of S2 is
therefore comparable to that of Omega Centauri, but $\sim$200 times more massive;  S2
is therefore unlikely to be a huge globular cluster given the predictions of
some current models (Schulz et al.\citealt{schulz15}). It does mean, however, that the BH
mass  (as estimated by the Greene \& Ho formalism) is 13\% of the mass of the S2 region, which is
significantly above the standard $M_\text{BH}$-to-bulge mass ratio
(Haering\& Rix\citealt{haering04}),
implying significant tidal stripping of the S2 mass---and that of S1 and N-G04---during merging.
S1 and N-G04 host black holes with masses that are four to eight times higher than that of S2.
Stellar velocity dispersions
are systematically higher in mergers than in isolated AGN host galaxies
(e.g., Liu et al.\citealt{liu12}, Stickley et al.\citealt{stickley14}, Medling et al.\citealt{medling15}).                                
Therefore, the black hole masses, based on the velocity dispersion, might be overestimated.

The masses we determined for the black holes in the nuclei N-G04, S-G04/S1, and S2 are on the same order as those
previously reported by other authors.
Based on SINFONI data
Engel et al.\cite{engel10} determine 
dynamical masses (Jeans modeling within r $<$ 250 pc each)
of the northern (N) and combined southern (S1+S2) nuclei based on the CO absorption
bandhead: 2.5 $\times 10^{9} M_{\odot}$ (N)
and 1.3 $\times 10^{10} M_{\odot}$ (S1+S2).
Based on the stellar dispersion $\sigma_\text{star}$ 200 km s$^{-1}$ (N) and
220 km s$^{-1}$ (S) and using the $M_\text{BH}$--$\sigma_\text{star}$ 
relation of Tremaine
et al.\cite{tremaine02}, they estimate black hole masses of
1.4$\pm{0.4}$ $\times 10^{8} M_{\odot}$ (N) and
2.0$\pm{0.4}$ $\times 10^{8} M_{\odot}$ (S1+S2).
Medling et al.\cite{medling11} derive
a mass of the combined southern system (S1+S2) based on Keck II Laser guide star
adaptive optics using OSIRIS data of the K-band CO absorption bandheads
to trace stellar kinematics and Jeans modeling.
They determine an upper limit of 2.0$\pm{0.2}$ $\times 10^{9} M_{\odot}$
and a lower limit of  8.7$\pm{0.3}$ $\times 10^{8} M_{\odot}$.
However, their absolute positioning of the black hole between the nuclei S1
and S2 (their Figs. 5 and 6) is inconsistent as the southern MERLIN and VLBA radio source
is overlaid on the S1 nucleus (see Figs.~\ref{weilngc6240_map_V_SN003_clean+GaiaDR2_v4b_ed.pdf}
and \ref{weilngc6240_map_Sigma_SN003_clean+GaiaDR2_v4b_ed.pdf}).

We did not derive an orbital mass for the N-G04 and S1--S2 system as the velocity difference 
between N-G04 and the S1--S2 system is quite small.
A statistical formula like that of Petrosyan\cite{petrosyan83} would lead to large errors
in the mass determination.

 \subsection{Merging history of the southern nuclei}

\begin{figure*}
\centering
\includegraphics[width=10.cm,angle=0]{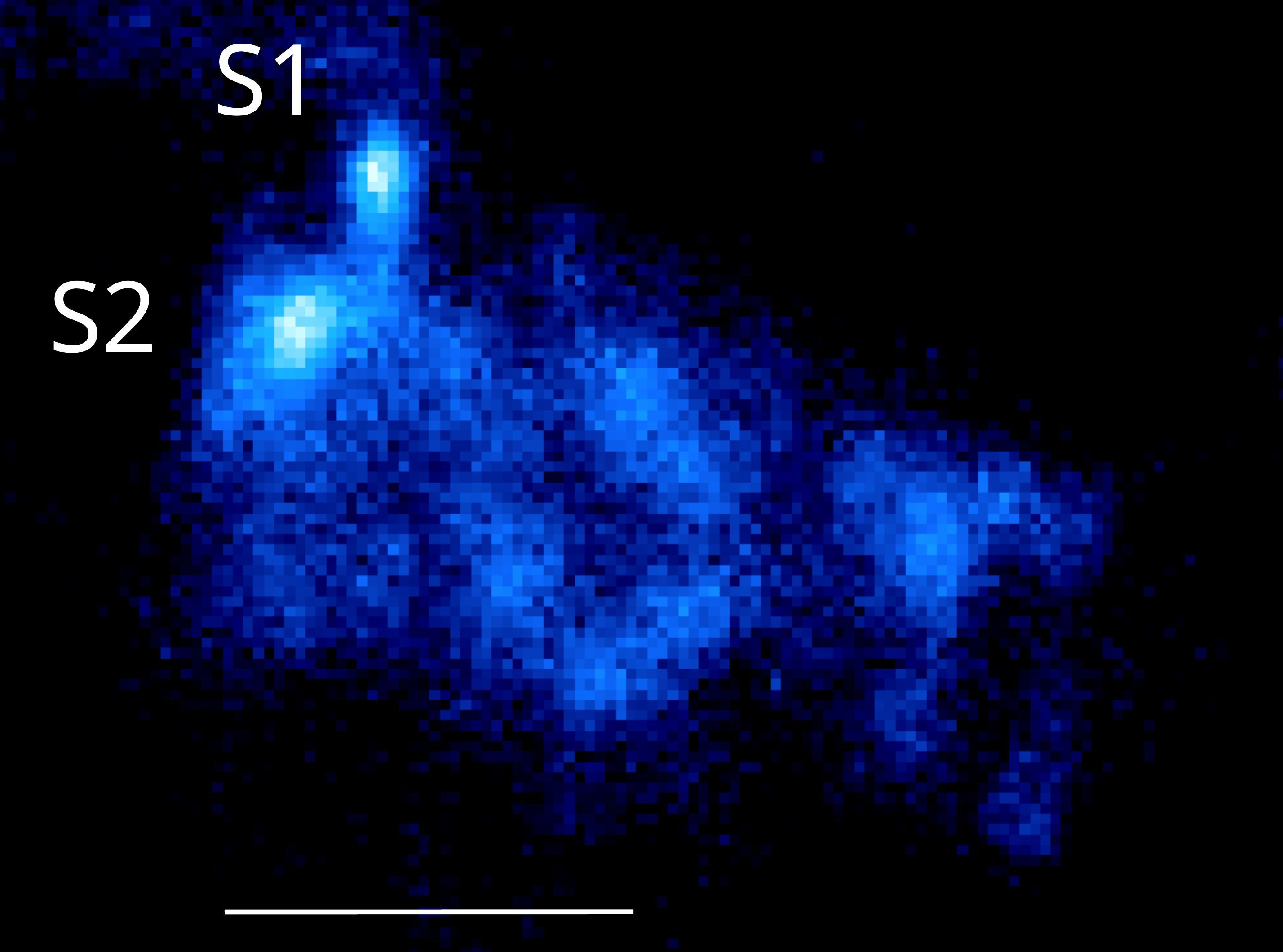}
\vspace*{2mm} 
\caption{ Enlarged H$\alpha$/[NII] image 
%(at $\lambda$6745\,\AA\ ) 
of the nuclei S1 and S2 and of the region 
southwest of the nuclei. North is to the top and east to the left. The white line is 1 arcsec long.
%Scale of the image: 2.95 x 2.65 arcsec, 1.39 x 1.25 kpc.  
}
\label{n6240_S1_S2_stream_nii_ochm.pdf}
\end{figure*}

A schematic view of the  general collision geometry in NGC\,6240 is presented
in Tecza et al. (\citealt{tecza00}). They assume that the southern
(double) nucleus is moving in the northeastern  direction  (their Fig. 11).
This is in accordance with the I-band contour levels,  caused by the old stellar component,
that are denser towards the northeastern direction than  the western direction
(see Figs.~\ref{weilngc6240_map_Ha6745+GaiaDR2_v4_large.pdf}, \ref{weilngc6240_map_I+GaiaDR2_v4_ed.pdf}).
Furthermore, this concept is supported by the complex extra H$\alpha$/[NII] emission component
southwest of the southern double nucleus
(see Fig.~\ref{n6240_S1_S2_stream_nii_ochm.pdf}).
The geometry of this extra H$\alpha$ emission component is reminiscent of the
Magellanic stream (e.g., Nidever et al. \citealt{nidever08}), which is composed
of two filaments. The two Magellanic stream filaments show periodic and
undulating spatial patterns and it is speculated that they are an imprint of the
LMC rotation curve. A similar picture has  evolved for the symbiotic
binary system R Aqr, where H$\alpha$ emission is arranged in an
elongated zig-zag pattern (Schmid et al.\citealt{schmid17}).
In our case the geometry of the H$\alpha$ filaments
might be caused by the orbit of the S1 and S2 components around each other
in combination with the general propagation of the  southern
double nucleus in the northeastern direction.
The rotational timescale  of S1 and S2  around each other (1 orbit)
amounts to $\sim$ 4 x 10$^{6}$ years,
based on the rotational velocity (v\,=\,158 [\kms{}]) of the nuclei and
on the S1 and S2 distance (198 pc).
We see about two and a half loops in Fig.~\ref{n6240_S1_S2_stream_nii_ochm.pdf}
corresponding to 1.25 orbits per nucleus under the simplified assumption
of circular motion. The same loop structure is seen not only  in H$\alpha$, but also 
in the emission lines of [OI], [NII], and [SII]
(see the [SII] structure in the Appendix (Fig.~\ref{n6240_S1_S2_stream_sii_ochm.pdf})).
Based on these loops we derive a look-back time of the past motion of
the S1 and S2 nuclei  from the southwest on the order of 5 x 10$^{6}$ years.
A more detailed simulation is beyond the scope of this manuscript. 

\subsection{Triple nucleus system in NGC\,6240}

The superposition of the optical images with the radio maps of
Gallimore et al.~\citep{gallimore04} shows that the northern active nucleus
N-G04 is centrally located in the northern bulge region, based on the CaII IR strength
of late-type giants. This position is neither identical with
the northern optical emission N at a distance of 0.33\,arcsec nor
with the IR emission `North 2' mentioned by Max et al.\cite{max05}.
The northern optical emission N is located at the edge of the bulge region
(see Figs.~\ref{weilngc6240_map_V_SN003_clean+GaiaDR2_v4b_ed.pdf},
~\ref{weilngc6240_map_Sigma_SN003_clean+GaiaDR2_v4b_ed.pdf}).
This emission of the N region might originate in the foreground as it shows the lowest absorption.

We propose that the southern components S1 and S2,  formerly thought to be one nucleus, are in fact two distinct nuclei. Specifically, S1 and S2 are two distinct entities 
in terms of

\begin{enumerate}[i.]
\item ionized gas structure (Fig.~\ref{weilngc6240_map_Ha6745+GaiaDR2_v4_large.pdf});
\item stellar structure (Fig.~\ref{weilngc6240_map_I+GaiaDR2_v4_ed.pdf});
\item spectral features in absorption and emission (Fig.~\ref{ochmall_multi.pdf});
\item their merging history (Fig.~\ref{n6240_S1_S2_stream_nii_ochm.pdf}).
\end{enumerate}
 More precisely, the H$\alpha$ redshifts of the gaseous components of S1 and S2 
 %($7163.\pm5.$\,km s$^{-1}$) 
 %($7473.\pm9$\,km s$^{-1}$)
 are comparable to the CaII IR absorption line redshifts  of S1  and S2 (Table~\ref{abs_vel}).
 %($7260.\pm29.$\,km s$^{-1}$)
 %($7576.\pm13.$\,km s$^{-1}$)
 The absorption components 
 associated with the nuclei arise in two distinct progenitor stellar bulges. These two central
 bulge regions S1 and S2 exhibit distinct
velocity dispersions.
%of 297.$\pm$29.\,km s$^{-1}$ (S1) and 180.$\pm$30.\,km s$^{-1}$ (S2).
Such a strong asymmetry in the dispersion
of a stellar component orbiting only one common single nucleus would be dispersed within
an orbiting timescale (Engel et al.~\citealt{engel10}). 
In addition, the intensity of the continuum and of the emission lines
in the intermediate region Sx is lower than that of S1 and S2. Furthermore,
the intrinsic absorption in Sx is lower in comparison to the
outer regions, and the recession velocity of the emission line gas in Sx is higher 
than in both
nuclei S1 and S2. Hence, we conclude that the two southern sources S1 and S2 are
two discrete nuclei separated by 198\,pc 
that rotate around each other.

\section{Conclusions}

We report the discovery of three nuclei in an advanced or final state of merging
within a region of only 1 kpc in NGC\,6240.
Thanks to MUSE we are able to show that
the formerly unresolved southern component actually consists of two nuclei
separated by only 198 pc. %representing the closest known distance between SMBHs.
The verification and detailed study of a system  with three nuclei, two 
of which are active and each with a mass in excess of $9\times10^{7} M_{\odot}$, is of great importance for the understanding of hierarchical galaxy formation 
via merging processes (Springel et al.\citealt{springel05})  since multiple mergers 
lead to a faster evolution of massive galaxies in comparison to
binary mergers.
So far it has been suggested that the formation of galactic nuclei with multiple SMBHs
is expected to be rare in the local universe (Kulkarni \& Loeb\citealt{kulkarni12}).

The superposition of the optical images of NGC\,6240 with radio maps
(Gallimore et al.~\citealt{gallimore04}) shows that the northern active nucleus
N-G04 is centrally located in the northern bulge region.
The southern nucleus S1 is active, based on its spatial matching with the 
compact radio source S-G04 (Gallimore et al.~\citealt{gallimore04}) 
and its associated hard X-ray emission. However, no radio counterpart
has been detected to date for the second southern nucleus S2,
suggesting that it is  an inactive nucleus.
Overall, the inner region of NGC\,6240 contains three nuclei in the final state of
merging. Two of them are active with respect to their radio emission.
Triple massive black hole systems might be of fundamental importance for the
coalescence of massive black hole binaries in less than a Hubble time
leading to the loudest sources of gravitational waves in the millihertz regime 
(Bonetti et al.~\citealt{bonetti18} and references therein).

\begin{acknowledgements}
This work has been supported by the DFG grant Ko 857/33-1. PMW was supported by BMBF Verbundforschung (project MUSE-NFM, grant
05A17BAA). AMI acknowledges support from the Spanish MINECO through project AYA2015-68217-P.  TC acknowledges support of the Agence Nationale de la Recherche (ANR) grant
FOGHAR (ANR-13-BS05- 0010), the OCEVU Labex (ANR-11-LABX-0060), and the
A'MIDEX project (ANR-11-IDEX-0001-02) funded by the 'Investissements d'avenir' French government program.
\end{acknowledgements}

\begin{appendix}
\section{Supplementary information}

\begin{table*}%[htbp]
    \centering
       \leavevmode
       \tabcolsep1.5mm 
\caption{
N component: Emission line centers, redshifts, line flux F in units of  10$^{-15}$\,erg\,s$^{-1}$\,cm$^{-2}$, and line widths (FWHM) of the strongest emission lines.
}
\begin{tabular}{lcccccc}
 \htopline
\hspace{3mm} Line & \mcc{$\lambda_\text{center}$} & \mcc{z} & \mcc{v} & \mcc{F}  & \mcc{FWHM} & \mcc{FWHM}\\
\hspace{3mm}      & \mcc{[\AA{}]}           &   & \mcc{[\kms{}]}&       & \mcc{[\AA{}]} &\mcc{[\kms{}]}\\
\hmidline   
\noalign{\smallskip}
\Nl{[N}{ii]}{6583}  &6739.8$\pm{}$0.2 &0.02376$\pm{}$.00003 &7126.5$\pm{}$9. &4271.$\pm{}$185. & 22.46$\pm{}$0.74 & 1000.$\pm{}$33.\\
H$\alpha$\,$\lambda${6563} &6718.5$\pm{}$0.1 &0.02372$\pm{}$.00002 &7117.4$\pm{}$5. &2012.$\pm{}$60. & 13.48$\pm{}$0.35 & 602.$\pm{}$16. \\
\Nl{[N}{ii]}{6548}  &6707.3$\pm{}$0.5 &0.02432$\pm{}$.00008 &7297.0$\pm{}$23. &1633.$\pm{}$57. & 22.46$\pm{}$0.60 & 1005.$\pm{}$25.\\
\Nl{[O}{i]}{6300}   &6453.6 $\pm{}$0.2 &0.02433$\pm{}$.00003 &7299.5$\pm{}$10. &699.$\pm{}$28. & 19.14$\pm{}$0.77 & 890.$\pm{}$36.  \\
\Nl{[O}{iii]}{5007} &5126.1$\pm{}$0.2 &0.02382$\pm{}$.00004 &7145.8$\pm{}$12. &390.$\pm{}$16. & 16.96$\pm{}$0.68  & 992.$\pm{}$38.\\
H$\beta$\,$\lambda${4861} &4975.0$\pm{}$0.3 &0.02338$\pm{}$.00004 &7014.0$\pm{}$18. &166.$\pm{}$8. & 10.06$\pm{}$0.40 & 606.$\pm{}$24.\\
\noalign{\smallskip}
\hbotline  
\end{tabular}
\label{N_emlines}
\end{table*}

\begin{table*}%[htbp]
    \centering
       \leavevmode
       \tabcolsep1.5mm 
\caption{
N-G04 component: Emission line centers, redshifts, line flux F in units of  10$^{-15}$\,erg\,s$^{-1}$\,cm$^{-2}$, and line widths (FWHM) of the strongest emission lines.
}
\begin{tabular}{lcccccc}
 \htopline
\hspace{3mm} Line & \mcc{$\lambda_\text{center}$} & \mcc{z} & \mcc{v} & \mcc{F}  & \mcc{FWHM} & \mcc{FWHM}\\
\hspace{3mm}      & \mcc{[\AA{}]}           &   & \mcc{[\kms{}]}&       & \mcc{[\AA{}]} &\mcc{[\kms{}]}\\
\hmidline   
\noalign{\smallskip}
\Nl{[N}{ii]}{6583}  &6746.9$\pm{}$0.2 &0.02483$\pm{}$.00003 &7450.0$\pm{}$9. &933.3$\pm{}$37. & 16.96$\pm{}$0.80 & 754.$\pm{}$30.\\
H$\alpha$\,$\lambda${6563} &6726.3$\pm{}$0.1 &0.02491$\pm{}$.00002 &7473.9$\pm{}$5. &806.2$\pm{}$24. & 15.18$\pm{}$0.46 & 677.$\pm{}$20. \\
\Nl{[N}{ii]}{6548}  &6711.4$\pm{}$0.5 &0.02495$\pm{}$.00007 &7484.8$\pm{}$23. &360.3$\pm{}$14. & 16.96$\pm{}$0.68 & 758.$\pm{}$30.\\
\Nl{[O}{i]}{6300}   &6458.3 $\pm{}$0.2 &0.02508$\pm{}$.00003 &7523.5$\pm{}$9. &239.8$\pm{}$11. & 16.51$\pm{}$0.83 & 767.$\pm{}$38. \\
\Nl{[O}{iii]}{5007} &5132.4$\pm{}$0.2 &0.02508$\pm{}$.00004 &7523.3$\pm{}$12. &77.0$\pm{}$14. & 10.65$\pm{}$0.53  & 623.$\pm{}$31.\\
H$\beta$\,$\lambda${4861} &4981.1:$\pm{}$1.0 &0.02464$\pm{}$.00018 &7390.6$\pm{}$60. &15.:$\pm{}$10. & 6.:$\pm{}$5. & 361.:$\pm{}$301.\\
\noalign{\smallskip}
\hbotline  
\end{tabular}
\label{N_emlines}
\end{table*}

\begin{table*}%[htbp]
    \centering
       \leavevmode
       \tabcolsep1.5mm 
\caption{
S-G04 component: Emission line centers, redshifts, line flux F in units of  10$^{-15}$\,erg\,s$^{-1}$\,cm$^{-2}$, and line widths (FWHM) of the strongest emission lines.  
}
\begin{tabular}{lcccccc}
 \htopline
\hspace{3mm} Line & \mcc{$\lambda_\text{center}$} & \mcc{z} & \mcc{v} & \mcc{F}  & \mcc{FWHM} & \mcc{FWHM}\\
\hspace{3mm}      & \mcc{[\AA{}]}           &   & \mcc{[\kms{}]}&       & \mcc{[\AA{}]} &\mcc{[\kms{}]}\\
\hmidline   
\noalign{\smallskip}
\Nl{[N}{ii]}{6583}  &6739.8$\pm{}$0.2 &0.02376$\pm{}$0.00003 &7126.5$\pm{}$9.       &1927.$\pm{}$67. & 13.93$\pm{}$0.49 & 620.0$\pm{}$22. \\
H$\alpha$\,$\lambda${6563} &6719.8$\pm{}$0.1 &0.02392$\pm{}$0.00002 &7176.8$\pm{}$5. &1699.$\pm{}$51. & 12.94$\pm{}$0.42 & 577.7$\pm{}$17. \\
\Nl{[N}{ii]}{6548}  &6704.8$\pm{}$0.4 &0.02390$\pm{}$0.00010 &7169.8$\pm{}$31.       & 739.4$\pm{}$26. & 13.93$\pm{}$0.49 & 623.3$\pm{}$22.\\
\Nl{[O}{i]}{6300}   &6449.1 $\pm{}$0.3 &0.02362$\pm{}$0.00004 &7085.2$\pm{}$14.      & 300.2$\pm{}$12. & 14.36$\pm{}$0.57 & 668.0$\pm{}$27.\\
\Nl{[O}{iii]}{5007} &5126.0$\pm{}$0.5 &0.02380$\pm{}$0.00010 &7139.8$\pm{}$30.        & 88.0 $\pm{}$5. & 9.81$\pm{}$0.44  & 574.1$\pm{}$26.\\
H$\beta$\,$\lambda${4861} &4979.0:$\pm{}$1.0 &0.02420$\pm{}$0.00021 &7261.0$\pm{}$62. & 23.3: $\pm{}$5. & 8.45:$\pm{}$1.8  & 509.:$\pm{}$107.\\
\noalign{\smallskip}
\hbotline  
\end{tabular}
\label{NG_emlines}
\end{table*}

\begin{table*}%[htbp]
    \centering
       \leavevmode
       \tabcolsep1.5mm 
\caption{
S1 component: Emission line centers, redshifts, line flux F in units of  10$^{-15}$\,erg\,s$^{-1}$\,cm$^{-2}$, and line widths (FWHM) of the strongest emission lines.
}

\begin{tabular}{lcccccc}
 \htopline
\hspace{3mm} Line & \mcc{$\lambda_\text{center}$} & \mcc{z} & \mcc{v} & \mcc{F}  & \mcc{FWHM} & \mcc{FWHM}\\
\hspace{3mm}      & \mcc{[\AA{}]}           &   & \mcc{[\kms{}]}&       & \mcc{[\AA{}]} &\mcc{[\kms{}]}\\
\hmidline   
\noalign{\smallskip}
\Nl{[N}{ii]}{6583}  &6739.3$\pm{}$0.2 &0.02368$\pm{}$0.00003 &7103.8$\pm{}$9. &2003.$\pm{}$78. & 14.6$\pm{}$0.57 & 650.$\pm{}$25.  \\
H$\alpha$\,$\lambda${6563} &6719.5$\pm{}$0.1 &0.02388$\pm{}$0.00002 &7163.1$\pm{}$9. &1797.$\pm{}$56. & 13.56$\pm{}$0.57 & 605.$\pm{}$25. \\
\Nl{[N}{ii]}{6548}  &6703.8$\pm{}$0.4 &0.02379$\pm{}$0.00010 &7137.0$\pm{}$31. &743.3$\pm{}$35. & 14.6$\pm{}$0.57 & 653.$\pm{}$25.\\
\Nl{[O}{i]}{6300}   &6448.1 $\pm{}$0.3 &0.02346$\pm{}$0.00004 &7037.6$\pm{}$14.  &266.5$\pm{}$11. & 13.5$\pm{}$0.54 &  626.$\pm{}$25. \\
\Nl{[O}{iii]}{5007} &5125.7$\pm{}$0.5 &0.02374$\pm{}$0.00010 &7121.9$\pm{}$30. &90.$\pm{}$5. & 10.9$\pm{}$0.45  & 639.$\pm{}$26.\\
H$\beta$\,$\lambda${4861} &4974.3$\pm{}$1.0 &0.02324$\pm{}$0.00021 &6971.0$\pm{}$62. &28.$\pm{}$3. & 8.42$\pm{}$1.0 & 508.$\pm{}$60.\\
\noalign{\smallskip}
\hbotline  
\end{tabular}
\label{S1_emlines}
\end{table*}

\begin{table*}%[htbp]
    \centering
       \leavevmode
       \tabcolsep1.5mm 
\caption{
Sx-clean component: Emission line centers, redshifts, line flux F in units of  10$^{-15}$\,erg\,s$^{-1}$\,cm$^{-2}$, and line widths (FWHM) of the strongest emission lines.
} 
\begin{tabular}{lcccccc}
 \htopline
\hspace{3mm} Line & \mcc{$\lambda_\text{center}$} & \mcc{z} & \mcc{v} & \mcc{F}  & \mcc{FWHM} & \mcc{FWHM}\\
\hspace{3mm}      & \mcc{[\AA{}]}           &   & \mcc{[\kms{}]}&       & \mcc{[\AA{}]} &\mcc{[\kms{}]}\\
\hmidline   
\noalign{\smallskip}
\Nl{[N}{ii]}{6583}  &6748.1$\pm{}$0.3 &0.02511$\pm{}$0.00005 &7504.8$\pm{}$14.       &730.9$\pm{}$29. & 13.36$\pm{}$0.53 & 594.$\pm{}$24.\\
H$\alpha$\,$\lambda${6563} &6727.5$\pm{}$0.2 &0.02510$\pm{}$0.00003 &7528.8$\pm{}$9. &679.8$\pm{}$27. & 12.40$\pm{}$0.50 & 553.$\pm{}$22.\\
\Nl{[N}{ii]}{6548}  &6710.5$\pm{}$0.5 &0.02481$\pm{}$0.00007 &7443.6$\pm{}$23.       &220.7$\pm{}$9. & 13.36$\pm{}$0.53 & 597.$\pm{}$24.\\
\Nl{[O}{i]}{6300}   &6455.5 $\pm{}$0.5 &0.02463$\pm{}$0.00008 &7389.9$\pm{}$24.     &124.6$\pm{}$6.  & 12.85$\pm{}$0.71  & 598.$\pm{}$30.\\
\Nl{[O}{iii]}{5007} &5131.2$\pm{}$1.0 &0.02484$\pm{}$0.00019 &7451.4$\pm{}$60.      & 63.5$\pm{}$16. & 15.68:$\pm{}$4. & 916.:$\pm{}$234.\\
H$\beta$\,$\lambda${4861} &4981.6$\pm{}$2.0 &0.02474$\pm{}$0.00040 &7421.5$\pm{}$123.&25.1$\pm{}$13.&  7.87:$\pm{}$4. & 474.:$\pm{}$241.\\
\noalign{\smallskip}
\hbotline  
\end{tabular}
\label{Sbetween_emlines}
\end{table*}

\begin{table*}%[htbp]
    \centering
       \leavevmode
       \tabcolsep1.5mm 
\caption{
S2 clean component: Emission line centers, redshifts, line flux F in units of  10$^{-15}$\,erg\,s$^{-1}$\,cm$^{-2}$, and line widths (FWHM) of the strongest emission lines.
} 
\begin{tabular}{lcccccc}
 \htopline
\hspace{3mm} Line & \mcc{$\lambda_\text{center}$} & \mcc{z} & \mcc{v} & \mcc{F}  & \mcc{FWHM} & \mcc{FWHM}\\
\hspace{3mm}      & \mcc{[\AA{}]}           &   & \mcc{[\kms{}]}&       & \mcc{[\AA{}]} &\mcc{[\kms{}]}\\
\hmidline   
\noalign{\smallskip}
\Nl{[N}{ii]}{6583}  &6746.4$\pm{}$0.3 &0.02476$\pm{}$0.00005 &7427.3$\pm{}$14.     &1063.$\pm{}$42.    & 15.94$\pm{}$0.64 & 709.$\pm{}$28. \\
H$\alpha$\,$\lambda${6563} &6726.3$\pm{}$0.2 &0.02491$\pm{}$0.00003 &7473.9$\pm{}$9. &907.8.$\pm{}$27. & 11.39$\pm{}$0.34 & 508.$\pm{}$15. \\
\Nl{[N}{ii]}{6548}  &6709.4$\pm{}$0.5 &0.02464$\pm{}$0.00007 &7393.2$\pm{}$23.      &469.6$\pm{}$19.   & 15.94$\pm{}$0.64 & 713.$\pm{}$28.\\
\Nl{[O}{i]}{6300}   &6452.2 $\pm{}$0.5 &0.02411$\pm{}$0.00008 &7232.8$\pm{}$24.      &209.3$\pm{}$10.  & 20.45$\pm{}$1.02 & 951.$\pm{}$47.\\
\Nl{[O}{iii]}{5007} &5131.0$\pm{}$1.0 &0.02480$\pm{}$0.00019  &7439.4$\pm{}$60.     &54.3$\pm{}$5.    & 10.42$\pm{}$1.04 & 609.$\pm{}$61.\\
H$\beta$\,$\lambda${4861} &4977.2$\pm{}$2.0 &0.02383$\pm{}$0.00041 &7150.2$\pm{}$124. &20.0:$\pm{}$4.   & 8.29:$\pm{}$1.37 & 500.:$\pm{}$101.\\
\noalign{\smallskip}
\hbotline  
\end{tabular}
\label{S2_emlines}
\end{table*}

\begin{figure*}
\centering
\includegraphics[width=10.cm,angle=0]{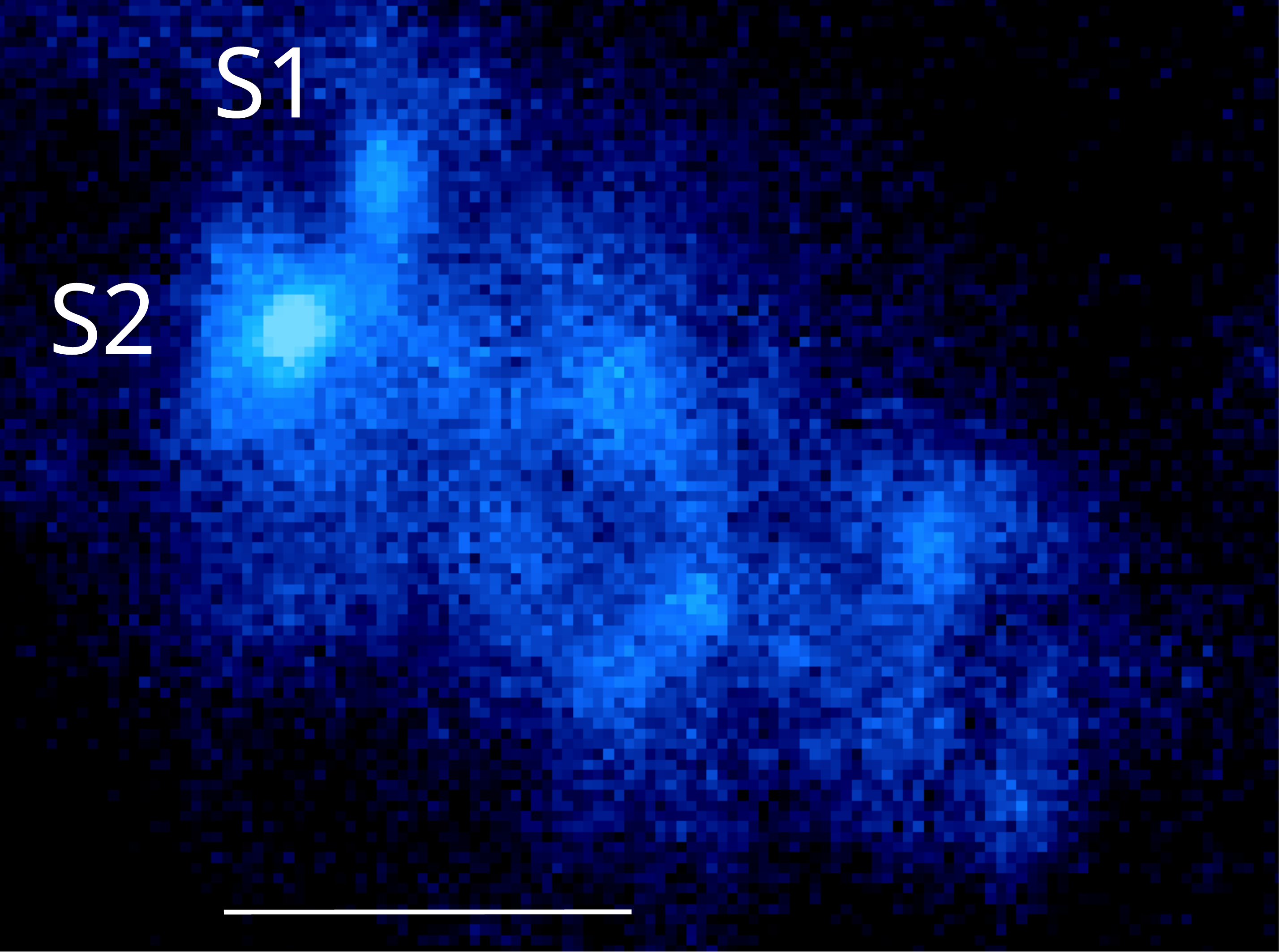}
\vspace*{2mm} 
\caption{ Enlarged [SII] image 
of the nuclei S1 and S2 and of the region 
southwest of the nuclei. North is to the top and east to the left. The white line is 1 arcsec long
(same as Fig.~\ref{n6240_S1_S2_stream_nii_ochm.pdf}).
}
\label{n6240_S1_S2_stream_sii_ochm.pdf}
\end{figure*}

\end{appendix}

%%%%%%%%%%%%%%%%%%%%%%%%%%%%%%%%%%%%%%%%%%%%%%%%%%%%%%%%%%%%%%%%%%%%%%%%%%%%5

\begin{thebibliography}{}
%
\bibitem[\protect\citeauthoryear{} {2008}]{allen08} Allen, M.G.,
 Groves, B.A., Dopita, M.A. et al. 2008, ApJS, 178, 20 
%
\bibitem[\protect\citeauthoryear{} {2010}]{bacon10} Bacon, R.,
 Accardo, M., Adjali, L., et al. 2010, SPIE Conference Series, Vol. 7735, 
Photo-Optical Instrumentation Engineers Conference Series, 8
%
  \bibitem[\protect\citeauthoryear{} {2014}]{bacon14} Bacon, R.,
 Vernet, J., Borosiva, E., et al. 2014, The Messenger , 157, 21 
%
\bibitem[\protect\citeauthoryear{} {1980}]{begelman80} Begelman, M.C.,
 Blandford, R.D., \&  Rees, M.J. 1980, Nature, 287, 307 
%
\bibitem[\protect\citeauthoryear{} {2018}]{bonetti18} Bonetti, M.,
 Haardt, F., Sesana, A., et al. 2018, MNRAS, 477, 3910 
%
\bibitem[\protect\citeauthoryear{} {2006}]{buchanan06}
 Buchanan, C.L., McGregor, P.J., Bicknell, G.V. et al. 2006, AJ, 132, 27
%
 \bibitem[\protect\citeauthoryear{} {2017}]{cappellari17}
 Cappellari, M. 2017, MNRAS, 466, 798C
%
\bibitem[\protect\citeauthoryear{} {2004}]{cappellari04}
 Cappellari, M., Emsellem, E. 2004, PASP, 116, 138
%
\bibitem[\protect\citeauthoryear{} {2013}]{dominguez13} Dominguez, A.,
 Siana, B., Henry, A.L. et al. 2013, ApJ, 763, 145 
%
\bibitem[\protect\citeauthoryear{} {1995}]{dopita95} Dopita, M.A.,
 \& Sutherland, R.S. 2013, ApJ, 455, 468 
%
\bibitem[\protect\citeauthoryear{} {1993}]{downes93} Downes, D.,
 Solomon, P.M. \& Radford, S.J.E. 1993, ApJ, 414, L13 
%
\bibitem[\protect\citeauthoryear{} {2010}]{engel10} Engel, H.,
 Davies, R.I., Genzel, R., et al. 2010, A\&A, 524, 56 
%
 \bibitem[\protect\citeauthoryear{} {1979}]{fosbury79} Fosbury, R.A. \&
 Wall, J.V. 1979, MNRAS, 189, 79 
%
 \bibitem[\protect\citeauthoryear{} {2004}]{gallimore04} Gallimore, J.F. \&
 Beswick, R. 2004, AJ, 127, 239 
%
\bibitem[\protect\citeauthoryear{} {2000}]{gebhardt00} Gebhardt, K.,
 Bender, R., Bower, G. et al. 2000, ApJ, 539, L13 
%
\bibitem[\protect\citeauthoryear{} {2000}]{genzel00} Genzel, R., \&
 Cesarsky, C.J. 2000, Ann. Rev. Astron. Astrophys. 38, 761 
%
\bibitem[\protect\citeauthoryear{} {2001}]{genzel01} Genzel, R., 
Tacconi, L.J., Rigopoulou, D. et al. 2001 ApJ, 563, 527
%
\bibitem[\protect\citeauthoryear{} {2019}]{goulding19} Goulding, A.D., 
Pardo, K., Greene, J.E. et al. 2019 ApJL, 879, L21
%
\bibitem[\protect\citeauthoryear{} {2006}]{greene06} Greene, J.E. \&
 Ho, L.C. 2006, ApJ, 641, L21 
%
\bibitem[\protect\citeauthoryear{} {2017}]{guerou17} Gu\'erou, A.,
   Krajnovic, D., Epinat, B. et al. 2017, A\&A, 608, 5G 
%
\bibitem[\protect\citeauthoryear{} {2004}]{haering04} Haering, N. \&
 Rix, H.-W. 2004, ApJ, 604, L89 
%
\bibitem[\protect\citeauthoryear{} {2011}]{hagiwara11} Hagiwara, Y..
Baan, W.A.\& Kloeckner, H.-R. 2011, AJ, 142, 17 
%
\bibitem[\protect\citeauthoryear{} {2003}]{ho03} Ho, L.C., Filippenko, A.V.,
Sargent, W.L.W. 2003 ApJ, 583, 159 
%
\bibitem[\protect\citeauthoryear{} {2007}]{hoffman07} Hoffman, L., Loeb, A.
2007 MNRAS, 377, 957 
%
\bibitem[\protect\citeauthoryear{} {2003}]{kauffmann03} Kauffmann, G.,
 Heckman, T.M., Tremonti, C. et al. 2003, MNRAS, 346, 1055 
%
\bibitem[\protect\citeauthoryear{} {2006}]{kewley06} Kewley, L.J., Growes, B.,
 Kauffmann, G. et al. 2006, MNRAS, 372, 961 
%
\bibitem[\protect\citeauthoryear{} {1984}]{kollatschny84} Kollatschny, W. \&
 Fricke, K. 1984, A\&A, 135, 171 
%
 \bibitem[\protect\citeauthoryear{} {1998}]{kollatschny98} Kollatschny, W. \&
 Kowatsch, P. 1998, A\&A, 336, L21 
%
\bibitem[\protect\citeauthoryear{} {2003}]{komossa03} Komossa, S.,
 Burwitz, V., Hasinger, G. et al. 2003, ApJ, 582, L15 
%
\bibitem[\protect\citeauthoryear{} {2013}]{kormendy13} Kormendy, J., Ho, L.,
 2013, ARA\&A, 51, 511 
%
\bibitem[\protect\citeauthoryear{} {2018}]{koss18} Koss, M.J., Blecha, L.,
 Bernhard, P. et al. 2018, Nature, 563, 214 
%
\bibitem[\protect\citeauthoryear{} {2013}]{kreckel13} Kreckel, K.,
 Groves, B., Schinnerer, E. et al. 2013, ApJ, 771, 62
%
\bibitem[\protect\citeauthoryear{} {2012}]{kulkarni12} Kulkarni, G., \&
 Loeb, A. 2012, MNRAS, 422, 1306 
%
\bibitem[\protect\citeauthoryear{} {2016}]{lindgren16} Lindegren, L., Lammers,
U., Bastrian, U. et al. 2016, A\&A, 595A, 4L
%
\bibitem[\protect\citeauthoryear{} {2012}]{liu12} Liu, X.,
 Shen, Y., Strauss, M.A. 20132, ApJ, 745, 94
%
\bibitem[\protect\citeauthoryear{} {2017}]{marziani17} Marziani, P., D'Onofrio,
M.D., Bettoni, D. et al. 2017, A\&A, 599, A83
%
\bibitem[\protect\citeauthoryear{} {2007}]{max07} Max, C.E., Canalizo, G.,
 de Vries, W.H. 2007, Science, 316, 1877 
%
\bibitem[\protect\citeauthoryear{} {2005}]{max05} Max, C.E., Canalizo, G.,
 Macintosh, B.A. et al. 2005, ApJ, 621, 738 
%
\bibitem[\protect\citeauthoryear{} {2012}]{mazzarella12} Mazzarella, J.M., Iwasawa, K.,
 Vavilkin, T. et al. 2012, AJ, 144, 125 
%
\bibitem[\protect\citeauthoryear{} {2011}]{medling11} Medling, A.M.,
 Ammons, S.M., Max, C.E. et al. 2011, ApJ, 743, 32 
%
\bibitem[\protect\citeauthoryear{} {2015}]{medling15} Medling, A.M.,
Vivian, U., Max, C.E. et al. 2015, ApJ, 803, 61 
%
\bibitem[\protect\citeauthoryear{} {2010}]{monreal10} Monreal-Ibero, A.,
 Arribas, S., Colina, L. et al. 2010, A\&A, 517, 28 
%
\bibitem[\protect\citeauthoryear{} {1987}]{netzer87} Netzer, H.,
     Kollatschny, W., Fricke, K. 1987, A\&A, 171, 41 
%
\bibitem[\protect\citeauthoryear{} {2008}]{nidever08} Nidever, D.L.,
 Majewski, S.R., Butler Burton, W. 2008, ApJ, 679, 432 
%
\bibitem[\protect\citeauthoryear{} {1983}]{petrosyan83} Petrosyan, A.R.
 1983, Sov.Astron.Lett., 9, 179 
%
\bibitem[\protect\citeauthoryear{} {2017}]{satyapal17} Satyapal, S., Secrest, N.J.,
     Ricci, C. et al. 2017 ApJ, 848, 126
% 
\bibitem[\protect\citeauthoryear{} {2017}]{schmid17} Schmid, H.M.,
     Bazzon, A., Milli, J. et al. 2017, A\&A, 602, A53 
%
\bibitem[\protect\citeauthoryear{} {2015}]{schulz15} Schulz, C.,
     Pflamm-Altenburg, J., Kroupa, P. 2015, A\&A, A93, 1 
%
\bibitem[\protect\citeauthoryear{} {2015}]{shetty15} Shetty, S. \&
 Cappellari, M. 2015, MNRAS, 454, 1332 
%
\bibitem[\protect\citeauthoryear{} {2005}]{springel05} Springel, V.,
 White, S.D.M., Jenkins, A. et al. 2005, Nature, 435, 629 
%
\bibitem[\protect\citeauthoryear{} {2014}]{stickley14} Stickley, N.R., Canalizo,
 G. 2014, ApJ, 786, 12 
%
\bibitem[\protect\citeauthoryear{} {2000}]{tecza00} Tecza, M., Genzel, R.,
 Tacconi, L.J. et al. 2000, ApJ, 537, 178 
%
\bibitem[\protect\citeauthoryear{} {2002}]{tremaine02} Tremaine, S., Gebhardt,
 K., Bender, R. et al. 2002, ApJ, 574, 740 
%
\bibitem[\protect\citeauthoryear{} {2004}]{valdes04} Valdes, F., Gupta, R.,
 Rose, J.A. et al. 2004, ApJS, 152, 251V 
%
\bibitem[\protect\citeauthoryear{} {2015}]{vasquez15} Vasquez, S., Zoccali, M.,
Hill, V. et al. 2015, A\&A, 580, A121 
%
\bibitem[\protect\citeauthoryear{} {1995}]{veilleux95} Veilleux , S.,
 Kim, S.-C., Sanders, D.B. 1995, ApJS, 98, 171 
%
\bibitem[\protect\citeauthoryear{} {2014}]{wang14} Wang , J.,
 Nardini, E., Fabbiano, G. 2014, ApJ, 781, 55 
%
%
\bibitem[\protect\citeauthoryear{} {2012}]{weilbacher12} Weilbacher, P.,
 Streicher, O., Urrutia, T., et al. 2012, SPIE Conference Series, Vol. 8451, 
Photo-Optical Instrumentation Engineers Conference Series, 0
%
  \bibitem[\protect\citeauthoryear{} {2014}]{weilbacher14} Weilbacher, P.,
 Streicher, O., Urrutia, T., et al. 2014, ASP Conference Series, 485, 451
%
\bibitem[\protect\citeauthoryear{} {2006}]{wright06} Wright, E.L., 2006, PASP, 118, 1711
%
\bibitem[\protect\citeauthoryear{} {1984}]{wright84} Wright, G. S., Joseph, R. D.,
     \& Meikle, W. P. S.1984, Nature, 309, 430 
%
\bibitem[\protect\citeauthoryear{} {2013}]{zhang13} Zhang, Z. T., Liang, Y. C.,
 Hammer, F. 2013, MNRAS, 430, 2605 
%
\end{thebibliography}
\end{document}